\documentclass[structabstract]{aa}
\usepackage{natbib}
\usepackage{graphicx}
\usepackage{txfonts}
\usepackage{array}
\begin{document}
\titlerunning{Multifrequency VLBA study of S5 0716+714, Paper II.}
\title{Multifrequency VLBA study of the blazar S5 0716+714 during the active state in 2004}
%  \thanks{Research supported ...}\fnmsep
 % \thanks{
%
 \subtitle{II. Large-scale jet kinematics and the comparison of the different methods of VLBI data imaging as applied to kinematic studies of AGN.}

 \author{E. A. Rastorgueva\inst{1}  \and K. J. Wiik\inst{1} \and A. T. Bajkova\thanks{Owing to different rules of Russian transcription in different journals, one can find other versions of this author's name: Baikova and Baykova}\inst{2} \and E. Valtaoja\inst{1} \and L. O. Takalo\inst{1} \and Y. N. Vetukhnovskaya\inst{3}  \and M. Mahmud\inst{4} }

\offprints{E. Rastorgueva, \email{eliras@utu.fi} }

\institute{Department of Physics and Astronomy, University of Turku, Tuorla Observatory, \\
           V\"{a}is\"{a}l\"{a}ntie 20, 21500, Piikki\"{o}, Finland \\
    \email{eliras@utu.fi, takalo@utu.fi}
  \and
        Central (Pulkovo) Astronomical Observatory, Russian Academy of Sciences, \\
             65-1, Pulkovskoe Shosse., St. Petersburg, 196140, Russia \\
    \email{bajkova@gao.spb.ru}
  \and
        Pushchino Radio Astronomy Observatory,  Lebedev Physical Institute, \\
             53, Leninskii Prosp., Moscow, 119991, Russia \\
    \email{yuvet@sci.lebedev.ru}
  \and
    Joint Institute for VLBI in Europe, \\
        4, Oude Hoogeveensedijk, Dwingeloo, 7991 PD, The Netherlands  \\
    \email{mahmud@jive.nl}
         }

\date{Received ... / Accepted ... }

% \abstract{}{}{}{}{}
% 5 {} token are mandatory

\abstract
{We study the jet kinematics of the blazar S5 0716+714 during its active state in 2003-2004
with multi-epoch VLBI observations.}
{We present a kinematic analysis of the large-scale (0-12 mas) jet of 0716+714,
based on the results of six epochs of VLBA monitoring at 5 GHz. Additionally, we compare kinematic results
obtained with two imaging methods based on different deconvolution algorithms.}
{The blazar 0716+714 has a diffuse large-scale jet, which is very faint compared with the bright compact core.
Experiments with simulated data showed that the conventional data reduction procedure based on the CLEAN 
deconvolution algorithm does not perform well in restoring this type of structure. This might be the reason why previous kinematic studies of this source yielded ambiguous results. In order to obtain accurate kinematics of this source, we independently applied two imaging techniques to the raw data: the conventional method, based on difference mapping, which uses CLEAN deconvolution, and the generalized maximum entropy method (GMEM) realized in the VLBImager package developed at the Pulkovo Observatory in Russia.}
{The results of both methods give us a consistent kinematic scenario: the large-scale jet of 0716+714 is diffuse and stationary. Differences between the inner (0-1~mas) and outer (1-12~mas) regions of the jet in brightness and velocity of the components could be explained by the bending of the jet, which causes the angle between the jet direction and the line of sight to change from $\sim 5^{\circ}$ to $\sim 11^{\circ}$.} 
{We tested the performance of the two imaging methods on real data and found that they yield similar kinematic results, but determination of the jet component positions by the conventional method was less precise. The method based on the GMEM algorithm is suitable for kinematic studies. It is especially effective for dim diffuse sources with the average brightness of several mJy with bright point-like features. For the source 0716+714, both methods worked at the limit of their capability.}

 \maketitle
 %
 %_____________________________________________________________________________

\section{Introduction}

The blazar S5 0716+714 at redshift z=0.31 \citep{Nilsson_z2008} is one of the most active sources of its
class. It is highly variable on timescales from hours to months across
all observed wavelengths. Intra-Day
Variability (IDV) has been detected in this blazar \citep{wagner1996, stalin2006,
montagni2006, gupta2009}. The IDV in optical and radio bands are reported to be correlated 
\citep{wagner1996}, which makes it the only candidate for a source
with an intrinsic origin of fast variation, which implies the presence of very small emitting regions within the source. At larger time scales this source has frequent flares in both radio and optical bands. Amplitudes of the flares increased abruptly after the year 2003  \citep[e.g.][and references
therein]{raiteri2003, nesci2005, bach2007, fuhrmann2008}. In 2007 and 2008 it
was also detected at very high-energy gamma rays with the MAGIC telescope \citep{MAGIC2009, MAGIC2009a}.

At the milliarcsecond scale (VLBI observations) 0716+714 has a core-dominated structure
with a bright point-like core and a faint diffuse jet pointing at the position angle of $\sim20^{\circ}$. 
At the scale of arcseconds (VLA observations), the source structure is also core-dominated, its jet is double-sided, located along the position angle of $\sim$~$100^{\circ}$ (misaligned with 
the milliarcsecond scale jet by  $\sim$~$80^{\circ}$), and has extended diffuse lobes \citep{wagner1996}.

The long-term kinematics of this source have been extensively studied, and two mutually exclusive models
have been suggested: one implies relatively fast outward motion of the superluminal components in the
jet with speeds ranging from 5\textit{c} to 16\textit{c} \citep[``fast scenario'',][]{bach2005},
 and the other claims the jet to be stationary, with significant non-radial (``precession'') motion of
the jet as a whole \citep[``stationary scenario''][]{britzen2009}. Both these scenarios are
based on the reanalysis of historical VLBI observations that span the years
1992-2001 (``fast'') and 1992-2006 (``stationary'') at frequencies from 5 to 43 GHz. 

Our analysis of the inner jet kinematics of 0716+714 at frequencies of 22, 43, and 86 GHz also revealed a fast-moving jet with apparent speeds from 8\textit{c} to 19\textit{c} on the spatial scale from 0 to 1.5~mas \citep{rastorgueva2008}. Here we present results of our short-term kinematic study at 5~GHz. The data set
covers six months, and the source was observed nearly every month. With this dense
data sampling we are able to study the fast structural changes in the jet of 0716+714.

The standard method of VLBI data reduction, which is used by the vast majority of
researchers in the field, includes the \textit{Difmap} package \citep{difmap1997} as an imaging facility,
which became conventional because of its simplicity and accessibility.
However, \textit{Difmap} has its downsides, and the main disadvantage of
this package, which uses the standard CLEAN algorithm by \citet{Hogbom1974} \citep{difmap1994},
is a problem with the reconstruction of extended emission \citep{cornwell1983}. Tests performed on
the simulated data have shown that the standard CLEAN, and \textit{Difmap} in general,
are especially inaccurate in the reconstruction of the sources whose structure could be described as 
bright point-like knots embedded in a region with faint extended emission
\citep[e.g.][]{bajkova2007, BeautyContest2004}. The morphology of 0716+714 at lower
frequencies (5-1.6 GHz) is consistent with this description. Therefore, a conventional
approach to this source may have yielded erroneous results. In order to make sure
that the structure of 0716+714 is correctly reconstructed, we independently applied both the generalized maximum entropy method (GMEM) and the conventional CLEAN to the same data set.

This paper is arranged as follows: section \ref{Obsevations} %``Observation'' 
describes the observations, section \ref{Data reduction} %~``Data Reduction''
contains a description of the \textit{a-priori} data calibration and an introduction to the imaging methods.
In section \ref{Results} %~``Results'' 
we present and discuss the kinematic results. The final 
kinematic scenario, a comparison with previous studies, and possible explanations of the apparent speed differences 
are presented in section \ref{Discussion}. %~``Discussion'' . 
% Maps and tables showing the results of each method could be found in the section \ref{Appendix}.

\section{Observations} \label{Obsevations}
We observed 0716+714 with the Very Long Baseline Array (VLBA) five times during the year 2004. Each
observing session lasted nine hours and covered five
frequencies (86, 43, 22, 5 and 1.6 GHz) in dual polarization mode. 
Here we describe the results of the 5~GHz observations. The five epochs in our experiment were named in alphabetical order: A, Feb 10; B, May 03; C, Jun 18;
D, Jul 29 and E, Aug 29, 2004. Additionally, we used one epoch from the unpublished polarization observation project of 37 northern BL Lac objects from the Kuehr \& Schmidt sample
\citep{KuehrSchmidt} by D. Gabuzda and M. Mahmud, in our notation epoch M, Mar 22, 2004.
Epoch M fills the gap between A and B of our experiment, making the time
sampling almost uniform with an interval of approximately one month between observations.

\section{Data reduction} \label{Data reduction}
\subsection{Calibration}
The data were correlated at the Socorro VLBA correlator. A-priori
amplitude and phase calibration were made at Tuorla
Observatory with NRAO's Astronomical Image Processing System
(AIPS). \textit{AIPS}'s standard VLBAPROC procedures were used for the amplitude and phase calibrations, 
scans on 3C279 were used to calculate instrumental delay and phase residuals, and OJ287 was used as an instrumental polarization (D-term) calibrator. The subsequent procedure for our two methods is described below.

\subsection{Image reconstruction: the conventional method} \label{CLEAN}
The first method we used was the conventional method, which is a
combination of CLEAN deconvolution \citep{Hogbom1974} with self-calibration and uv-plane
Gaussian model fitting. This method is used by most of the astronomical community in 
studies of outflow structure and kinematics.

\begin{figure*}[ht]
\centering
    \includegraphics[angle=0,width=0.30\textwidth]{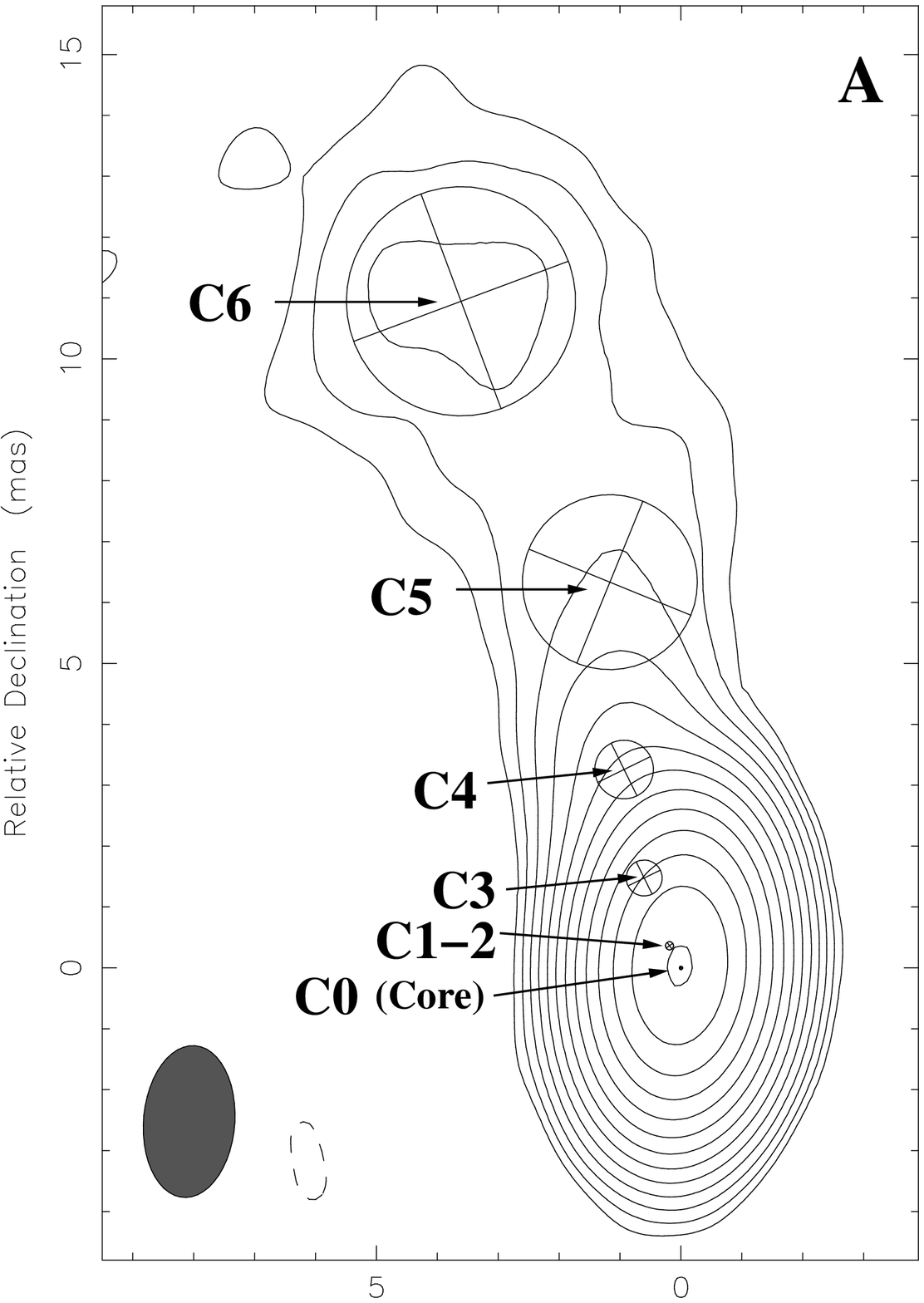}
\includegraphics[angle=0,width=0.287\textwidth]{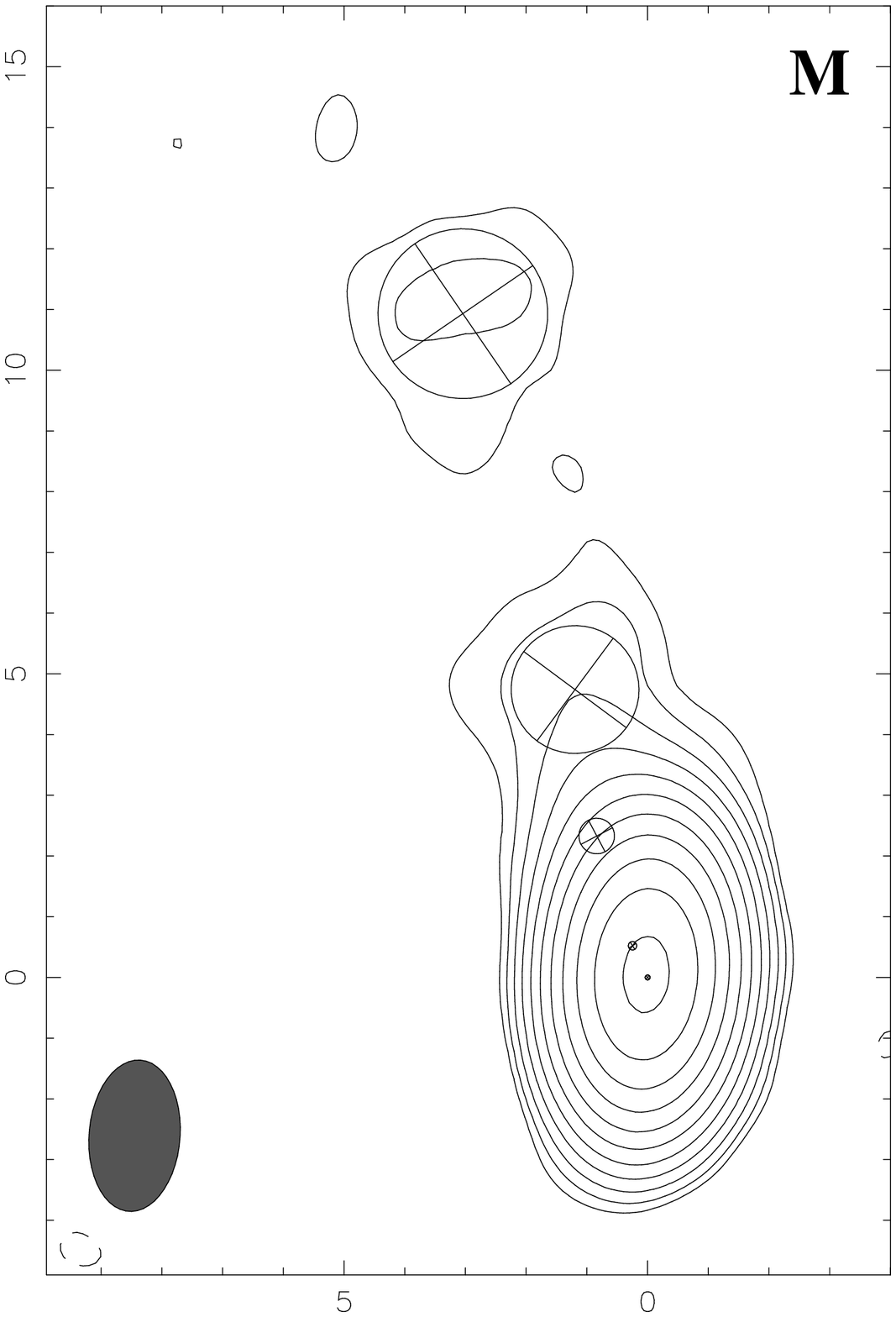}
\includegraphics*[angle=0,width=0.29\textwidth]{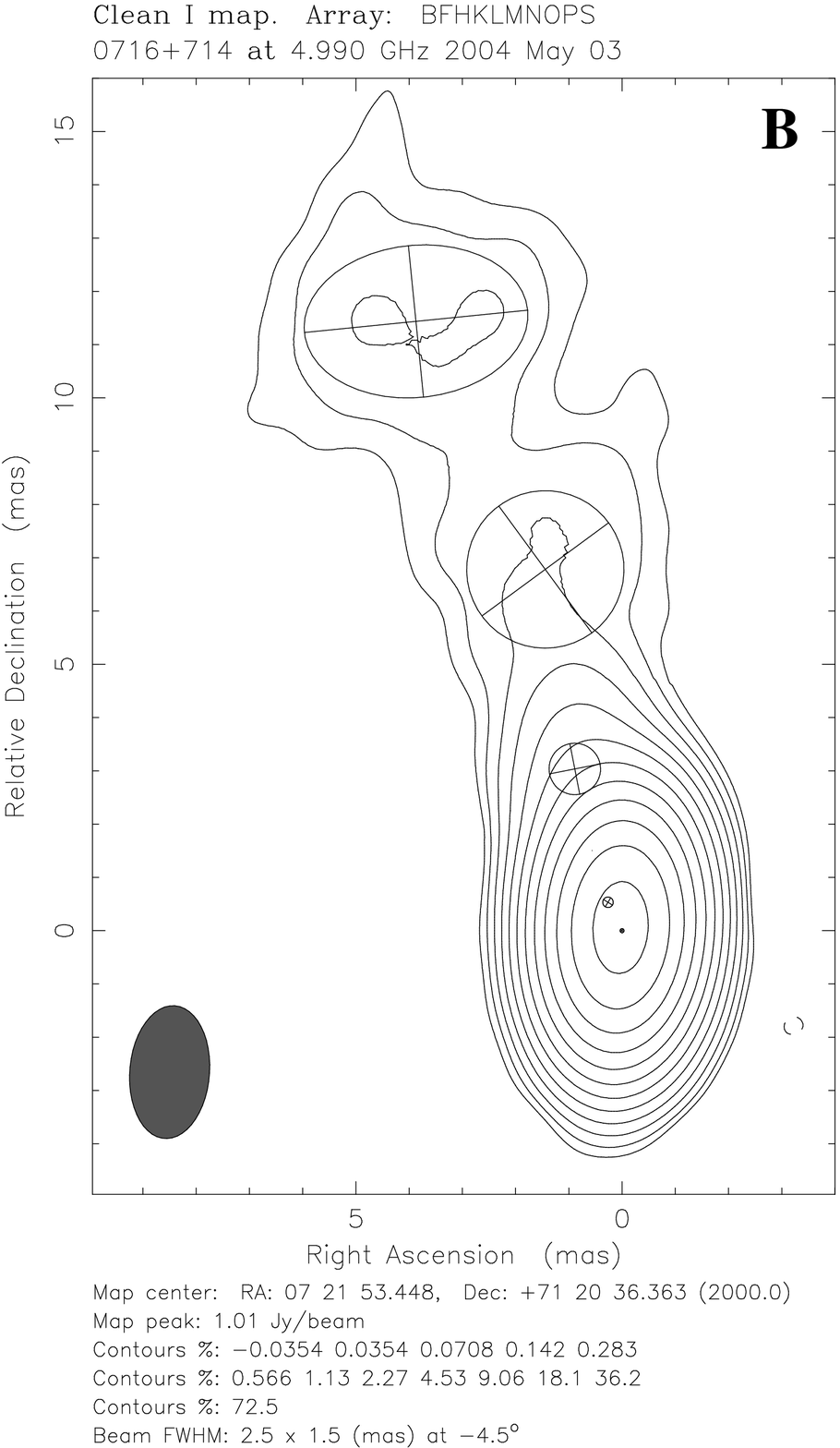}
\includegraphics[angle=0,width=0.30\textwidth]{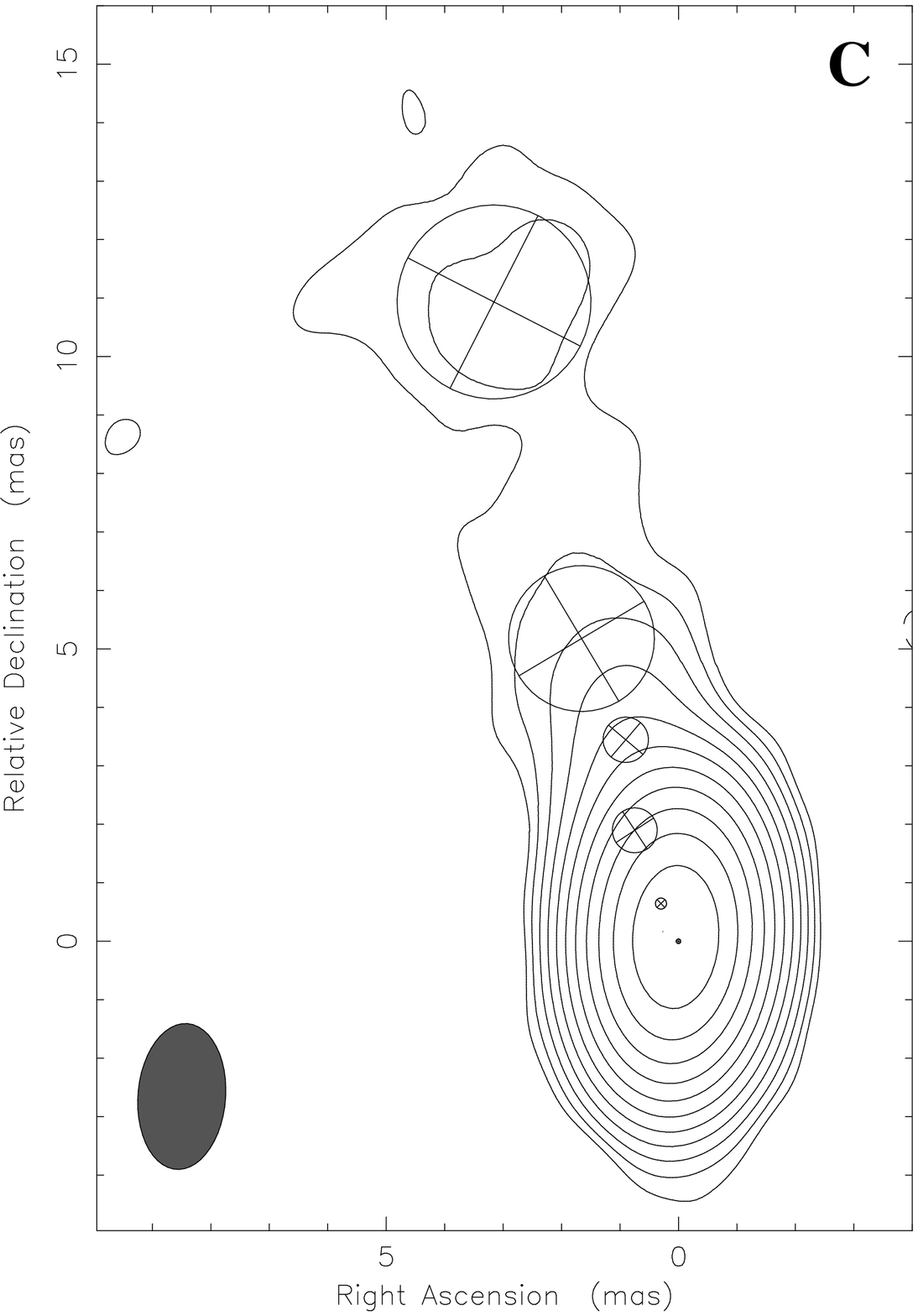}
\includegraphics[angle=0,width=0.283\textwidth]{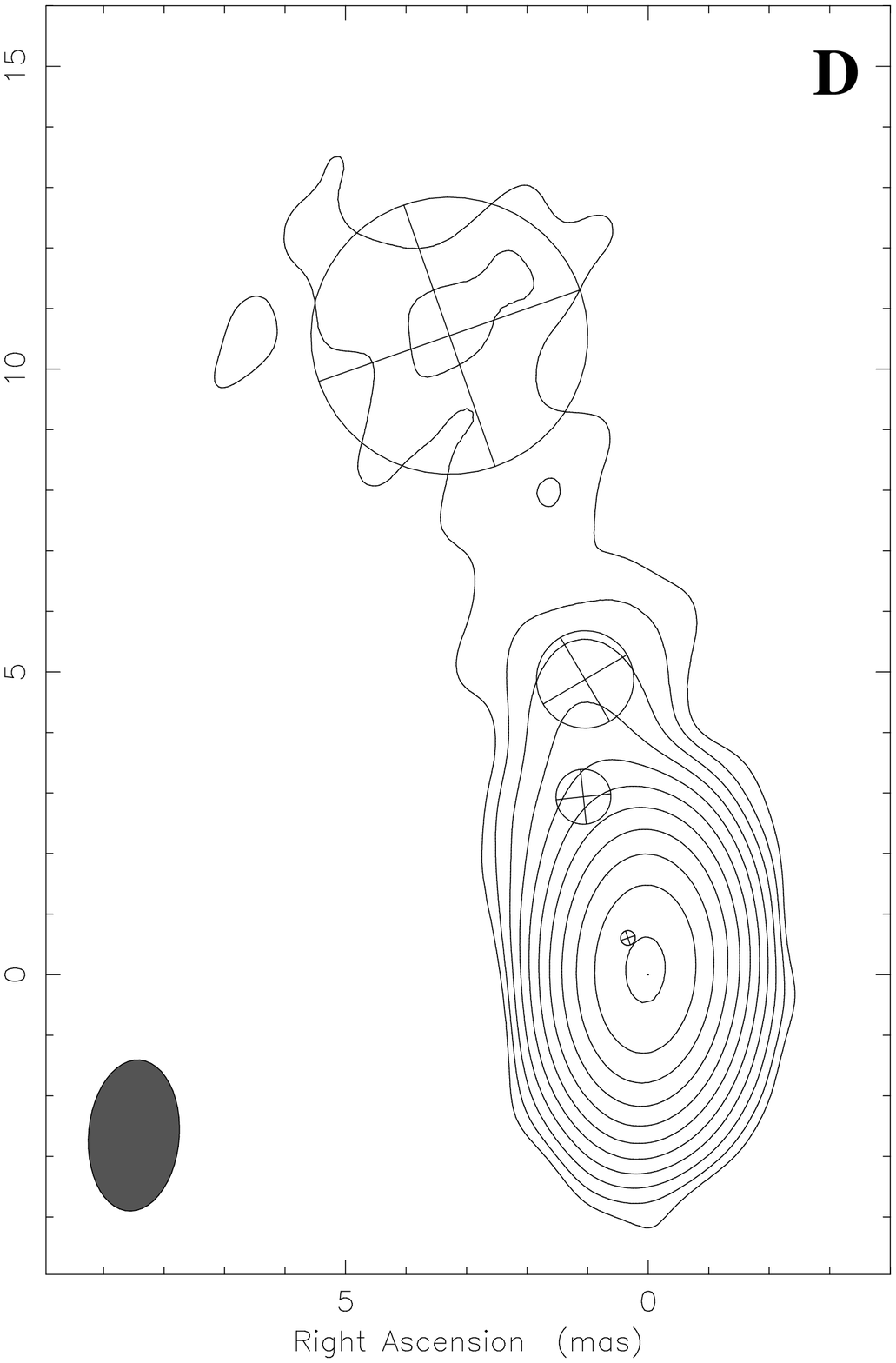}
\includegraphics[angle=0,width=0.283\textwidth]{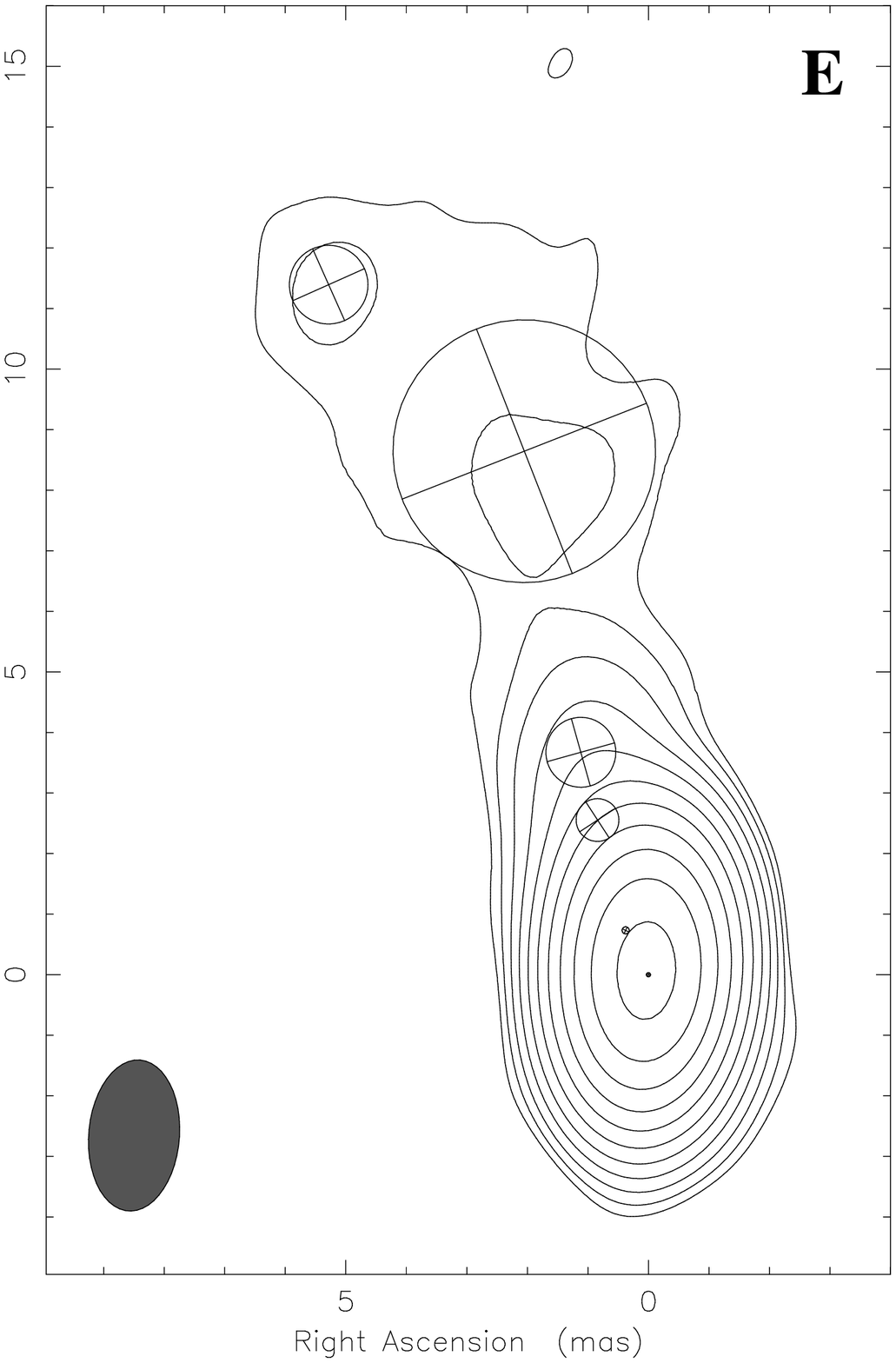}
    \caption{Images of 0716+714 obtained using the conventional method: CLEAN maps, convolved with the averaged beam of 1.5$\times$2.5 mas, PA $= -4.5^{\circ}$. Circles with crosses  are the 
{\em Modelfit} Gaussian components superimposed on the image. The Gaussian model fitting was performed in the uv-plane. Axes are right ascension and declination in milliarcseconds, measured with regard to the core.}
    \label{map:CLEANmaps}
\end{figure*}

We used the Caltech \textit{Difmap} package
\citep{difmap1997} for imaging, employing a CLEAN algorithm in a self-calibration loop: 
a model of the source, consisting of point-like clean-components 
is created with CLEAN, then the antenna gains are self-calibrated with the
model, and the procedure is repeated. Self-calibrations were performed in the following way:
 several phase self-calibration loops were followed by
an amplitude one, with the solution interval of
amplitude self-calibration gradually decreasing from the time of the whole observation
to the data averaging time (0.5~min). Finally, we performed amplitude self-calibration with the solution interval equal to the averaging time. Then the final clean-component	model was discarded, and a new source model was created. We used the \textit{Difmap} procedure, \textit{Modelfit}, to create a model consisting of circular components with a Gaussian distribution of flux, and fit it to the visibility data. Except for one case, only circular Gaussian components were used to parameterize the data in order to reduce the number of free parameters. For all models, the brightest optically thick component at the southernmost end of the jet was considered to be the core, as in the previous research. Positions of other components were measured with regard to this. The resulting maps with the superimposed model components are presented in Fig.~\ref{map:CLEANmaps}, %, \ref{Appendix}~Appendix, 
and map parameters for all epochs are listed in
Table~\ref{tb:CLEAN_map_parameters}.
\begin{table}[ht]
\caption[]{Map parameters for the conventional method.}\label{tb:CLEAN_map_parameters}
\centering
\begin{tabular}{ccccc}\hline
Epoch & $S_{tot}$ &   Beam                & $S_{peak}$ & $3\sigma$ \\
      & [Jy]      & mas$\times$ mas,[$^o$]& [Jy/beam]  & [mJy/beam] \\
\hline
2004.11 & 1.46 &$2.5\times$1.5,-4.5 & 1.41 & 0.33 \\
2004.22 & 1.13 &$2.5\times$1.5,-4.5 & 1.08 & 0.89 \\
2004.34 & 1.06 &$2.5\times$1.5,-4.5 & 1.01 & 0.36 \\
2004.46 & 1.16 &$2.5\times$1.5,-4.5 & 1.10 & 0.54 \\
2004.58 & 0.90 &$2.5\times$1.5,-4.5 & 0.84 & 0.72 \\
2004.66 & 0.87 &$2.5\times$1.5,-4.5 & 0.81 & 0.63 \\
\hline
\hline
\end{tabular}
\end{table}
%\subsubsection{\textit{Difmap}: CLEAN+Modelfit}
%Details of the observation setup of our experiment could be found
%in \citet{rastorgueva2008}.
The errors of the position and the flux estimates of the components were
calculated with the {\em Difwrap} package \citep{lovell2000difwrap}, where the parameter to be analyzed (position, flux or FWHM) is varied around the best-fit value in a grid of a given step size and textit{Modelfit} is run for each step. The result of this process is plotted as a $\chi^{2}$ distribution, and the user can visually inspect the model vs a uv-data plot or a residual map for each point of the grid. Visual inspection is based on the relative $\chi^{2}$ changes.
Because the number of degrees of freedom of the model are not exactly known,
this method of error estimate based on visual inspection yields more
reliable results than the standard $\chi^{2}$ analysis.
In our case it was not possible to use the model vs a visibility plot, %of correspondence between
%observed and model visibilities provided by {\em Difwrap},
because we dealt with very faint components of the jet: changes in the model visibilities between different grid points were
too small to be detected with the naked eye. Instead, we estimated the goodness of fit based on the appearance of the residual map and its changes from one grid point to another. We applied an intuitive criterion
proposed by \citet{savolainen2006a} to determine the maximal acceptable deviation
from the best-fit model: the difference between a model and observed data is considered
significant if the structure appearing in the residual map would look strong enough that 
it should be cleaned, if it had appeared during the imaging process.
Because the flux of such a feature changes gradually from one grid point to another, it may be
difficult to determine at which step the discrepancy between the
observed data and the model becomes large enough. In these cases we used an empirical
`2$\sigma$-rule,' which we worked out by analysis of both the imaging process and
the error estimate procedure: the flux is considered
`excessive' and should be `cleaned out' by placing a model component
in its location if the rms noise in the area under scrutiny is twice as high as the average rms noise
at the outskirts of the map. At the grid point where the noise level exceeds twice the rms, we
assumed that the parameter value had reached the limit. This deviation from the best-fit value was assumed to be the limiting error, $\Delta$. A more detailed
discussion of VLBI data error estimate with \textit{Difwrap} and
a comparison with the method of variance about the
best fit can be found in \citet{savolainen2006a}.

In most cases the  {\em Difwrap} package yields asymmetric
uncertainties for the model parameters, which leads to a bias
in the resulting models. This bias was corrected for with the method suggested by
\citet{agostini}, based on the probabilistic approach to the bias correction:
let us assume that the negative limiting error
of the component's separation from the core r is $\Delta_{neg}$r,
and the positive limiting error is $\Delta_{pos}$, and
$\Delta_{neg} \neq \Delta_{pos}$. We calculated the unbiased
limiting error as an arithmetic average of the two limiting errors:
$\Delta_{un}$=($\Delta_{neg}$+$\Delta_{pos}$)/2. The standard
deviation of r was calculated as
$\sigma_{r}$=$\Delta_{un}$/3. The unbiased value of the model parameter
(in this case, separation from the core) is not equal to the result of the 
model fitting and is $r_{un}$=r+($\Delta_{pos}$-$\Delta_{neg}$)/2.

\subsection{Image reconstruction: the maximum entropy method (MEM), an attempt}\label{cpt:MEM}

Another popular deconvolution algorithm in radio
interferometry is the maximum entropy method invented by Jaynes \citep{jaynes1957},
which was first applied to image reconstruction by Frieden \citep{frieden1972}. The
maximum entropy method is an extremely nonlinear method that
leads to maximally smoothed images subject to constraints. The
positivity of the sought-for solution forces super-resolution on
bright isolated objects.

There are many forms of entropy
\citep{narayan1986}. We used the \textit{AIPS} task \textit{VTESS}, which
uses the entropy functional in the form \citep{cornwell1995}

\begin{eqnarray}
&& E(I)= -\int I(x,y)\ln[I(x,y)/M(x,y)] dx dy~, \label{cornwell_entropy} \\
&& I(x,y)>0~, \nonumber
\end{eqnarray}
where $M(x,y)$ is a default image that represents the optimum
solution in the absence of data. Usually, the default image is
taken to be flat or a lower resolution image.

Given a form of entropy, the MEM defines the best image that
maximizes $E(I)$ subject to the data constraints: 1)~fit the
observed visibilities $V_k$ to within the error term $\epsilon_k$:
$\emph{F}(I)=V_k+\epsilon_k$; 2)~fit the total flux: $\int
I(x,y)dxdy=F_0$; 3)~the rms error must be plausible:
$\sum\frac{|\epsilon_k|^2}{\sigma_k^2}=\Omega$.

In \textit{AIPS}, the Cornwell-Evans algorithm
\citep{cornwellevans1985} is used for the realization of the MEM. This algorithm uses a simple Newton-Raphson
approach to optimize the functional (\ref{cornwell_entropy}) subject to the constraints upon the rms
and total power enforced by the Lagrange multipliers.

%Another method considered was AIPS implementation of the maximum-entropy method (MEM).
%We used AIPS task VTESS, based on the algorithm VM \citep[][and references therein]%%{cornwellevans1985},%
%developed by the NRAO.
We made an attempt to use the MEM method to recover the faint emission of the 0716+714 jet. 
We imported visibility data, self-calibrated in \textit{Difmap} (see Section \ref{CLEAN}), to \textit{AIPS} and used the task VTESS to create the final image. 

We used the task IMAGR to clean the central part of the image containing the bright core in order to 
avoid circular-shaped artifacts around the point-like source. Nevertheless, the image quality of this method was poor and did not allow us to correctly perform the kinematic analysis. Therefore, we conclude that the MEM method is not suitable for recovering the emission of the sources with bright point-like features and faint underlying diffuse emission.

\subsection{Image reconstruction: the generalized maximum entropy method (GMEM)}
The eneralized maximum entropy method has been developed and implemented in the Pulkovo VLBI data reduction software package \textit{VLBImager} by A.~Baikova \citep{bajkova2007}.
The mapping technique realized in this package utilizes a self-calibration algorithm \citep{cornwellformalont1999}
combined with the GMEM
\citep{friedenbajkova1994, bajkova2007}, which is used as a deconvolution procedure.

The GMEM is designed for the
reconstruction of sign-variable functions, therefore it allows one to
obtain unbiased solutions. The bias of the solution is one of the
problems of the conventional MEM \citep{cornwellbraunbriggs1999},
which may lead to a substantial nonlinear distortion of the
final image if the data contain errors \citep{bajkova1995}.

For the GMEM, the Shannon-entropy functional has the form

\begin{eqnarray}\label{shannon_entropy}
&& E(\alpha)= -\int (x^p(t) \ln(\alpha x^p(t)) + x^n(t) \ln(\alpha x^n(t)))~dt~,  \\
&&  x^p(t)>0~,~~~x^n(t)>0~, \nonumber
\end{eqnarray}
where $x^p(t)$ and $x^n(t)$  are the positive and negative
components of the sought-for image $x(t)$, i.e. the equation
$x(t)=x^p(t)-x^n(t)$ holds. $\alpha>0$ is a parameter
responsible for the accuracy of the separation of the negative and positive
components of the solution $x(t)$, and therefore
critical for the resulting image fidelity. It is easy to
show \citep{bajkova2007} that solutions for $x^p(t)$ and $x^n(t)$ 
obtained with the Lagrange optimization method are
connected by the expression $x^p(t) \cdot x^n(t) = \exp(-2-2\ln
\alpha) = K(\alpha)$, which depends only on the parameter
$\alpha$. This parameter is responsible for dividing the positive and negative
parts of the solution: the larger $\alpha$ is, the more accurate
is the discrimination. On the other hand, the value of $\alpha$ is constrained
by computational limitations. The main constraint comes from the
$\chi^{2}$ term in the optimized  functional, which depends
on the data errors. The larger a standard deviation is, the higher value of
$\alpha$ could be set. If data are very accurate, a lower value of
$\alpha$ is needed. In practice, $\alpha$ is chosen empirically.
In our case we had to compromise between data errors, which determine
resolution of the resulting MEM solution, and a need to divide the positive
and negative parts of the solution as accurately as possible.
It is fair to say that given fixed errors in the data, a maximum possible chosen value of $\alpha$
provides us with the best possible resolution of the MEM-solution. In this work we used $\alpha=100000$.
The algorithm takes into account errors of the measurements. It implies a solution of the following two-dimensional problem of conditional optimization, given in the discrete form

\begin{eqnarray}\label{anisa_functional}
\lefteqn{\min~~\{\sum_{k=1}^N\sum_{l=1}^N [x_{kl}^p\ln(\alpha x_{kl}^p)+
x_{kl}^n\ln(\alpha x_{kl}^n)]+} \\
&& ~~~~~~~~~~~~~~~~~~~~~~~~~~~~~~~~~~~~~~~~~~~~~~~~~~~~~+\sum_{m=1}^M\frac{(\eta_m^{re})^2+(\eta_m^{im})^2}{\sigma_m^2}\}~,  \nonumber \\
&& \sum_{k=1}^N\sum_{l=1}^N (x_{kl}^p-x_{kl}^n)a_{kl}^m+\eta_m^{re}=V_m^{re}~, \nonumber \\
&& \sum_{k=1}^N\sum_{l=1}^N (x_{kl}^p-x_{kl}^n)b_{kl}^m+\eta_m^{im}=V_m^{im}~, \nonumber
\end{eqnarray}
where \textit{k} and \textit{l} are the pixel numbers in a source map of the
size $N\times N$, $m=1,...,M$ is a number of the visibility
function measurement, $a_{kl}^m$ and $b_{kl}^m$ are the constant
coefficients corresponding to the Fourier-transform, $V_m^{re}$ and
$V_m^{im}$ are the real and imaginary parts of the complex
visibility function (the Fourier spectrum of a source) and $\eta_m^{re}$ and 
$\eta_m^{im}$ are the real and imaginary parts of the additive
visibility noise with normal distribution over zero value with
known dispersion $\sigma_m$.

As we can see from equation (\ref{anisa_functional}), the optimized functional
has two parts: a Shannon-entropy functional and a functional that
is an estimate of the difference between the reconstructed
visibility spectrum and the measured data according to a $\chi^2$ criterion.
This latter functional can be considered to be an additional stabilizing
term acting to provide a further regularization of the solution
above what is possible with the entropy functional alone. Solution
normalization, necessary for the MEM is obtained by limiting the total
flux of the source:

\begin{eqnarray}
&& \sum_{k=1}^N\sum_{l=1}^N (x_{kl}^p-x_{kl}^n) = F_0~. \nonumber
\end{eqnarray}

Numerical experiments show that the  GMEM combined with the difference 
mapping technique allows the restoration of the source structures consisting of
bright compact features embedded in a weaker extended background with maximal
accuracy \citep{bajkova2007}.

%\subsubsection{Model fitting with Gaussians}\label{GMEM_model_fitting_section}
%\subsubsection{Data processing}
\begin{figure*}[ht]
    \centering
        \includegraphics*[angle=0,width=0.31485\textwidth]{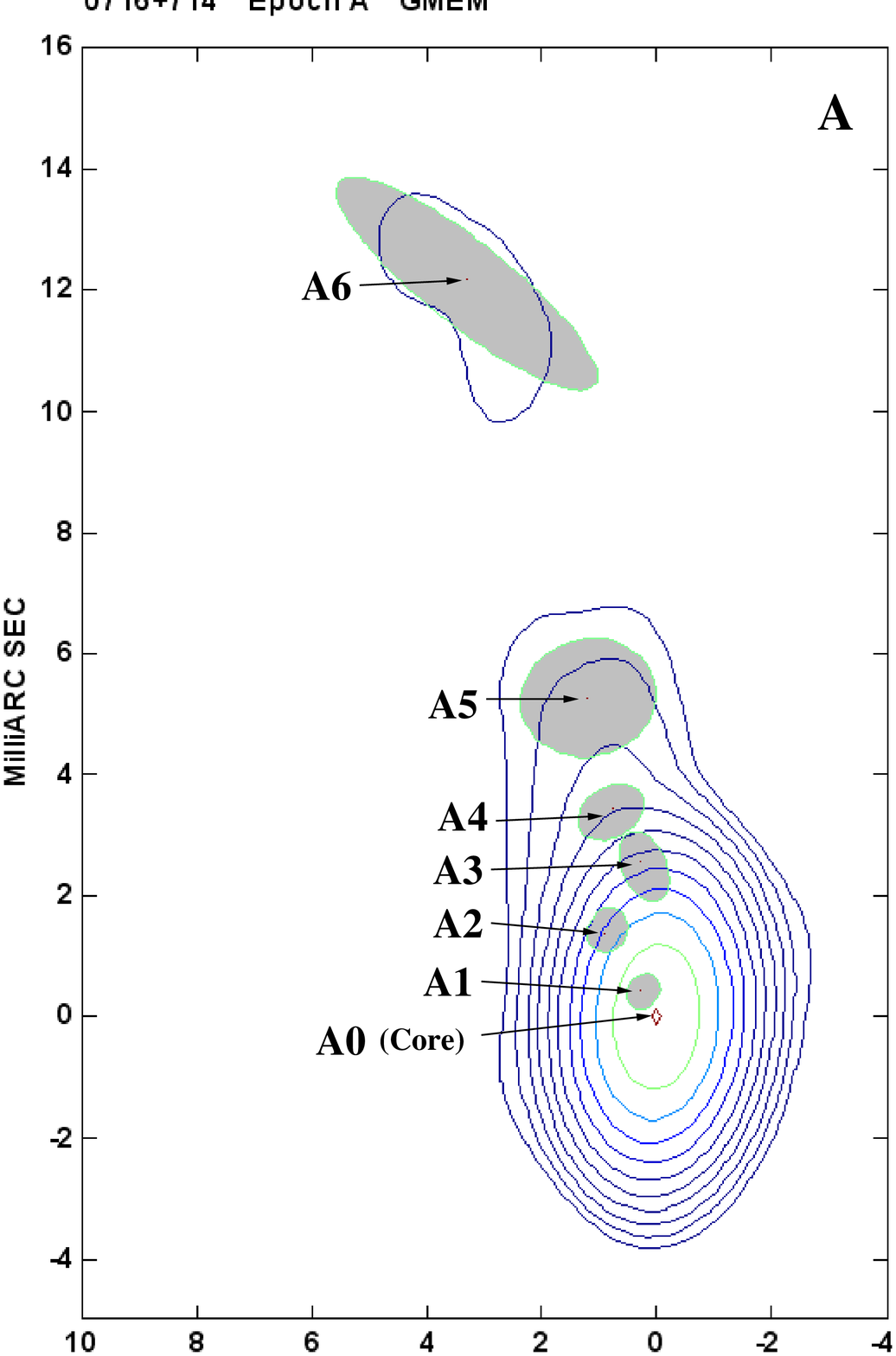}
\includegraphics*[angle=0,width=0.30\textwidth]{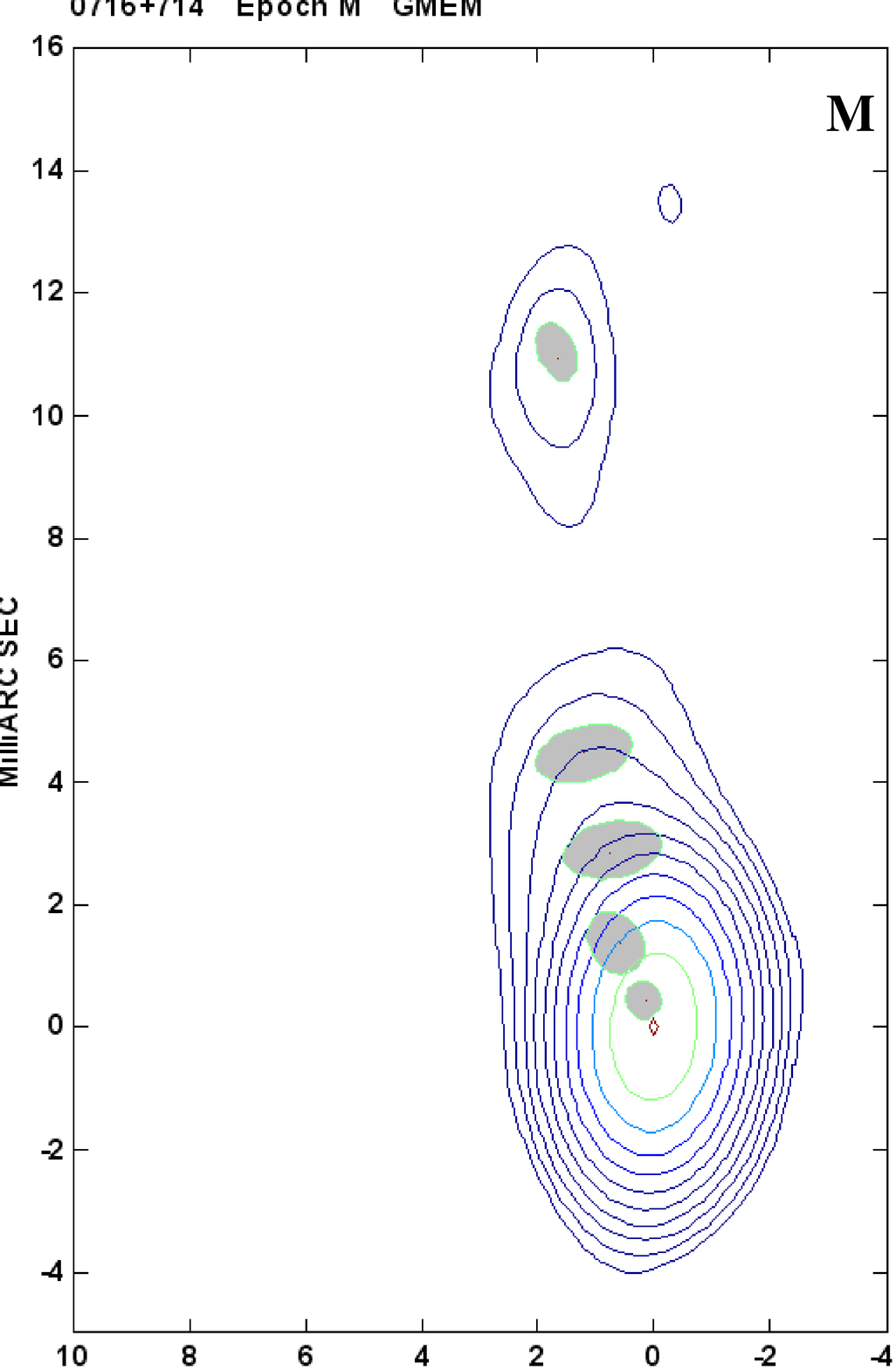}
\includegraphics*[angle=0,width=0.30\textwidth]{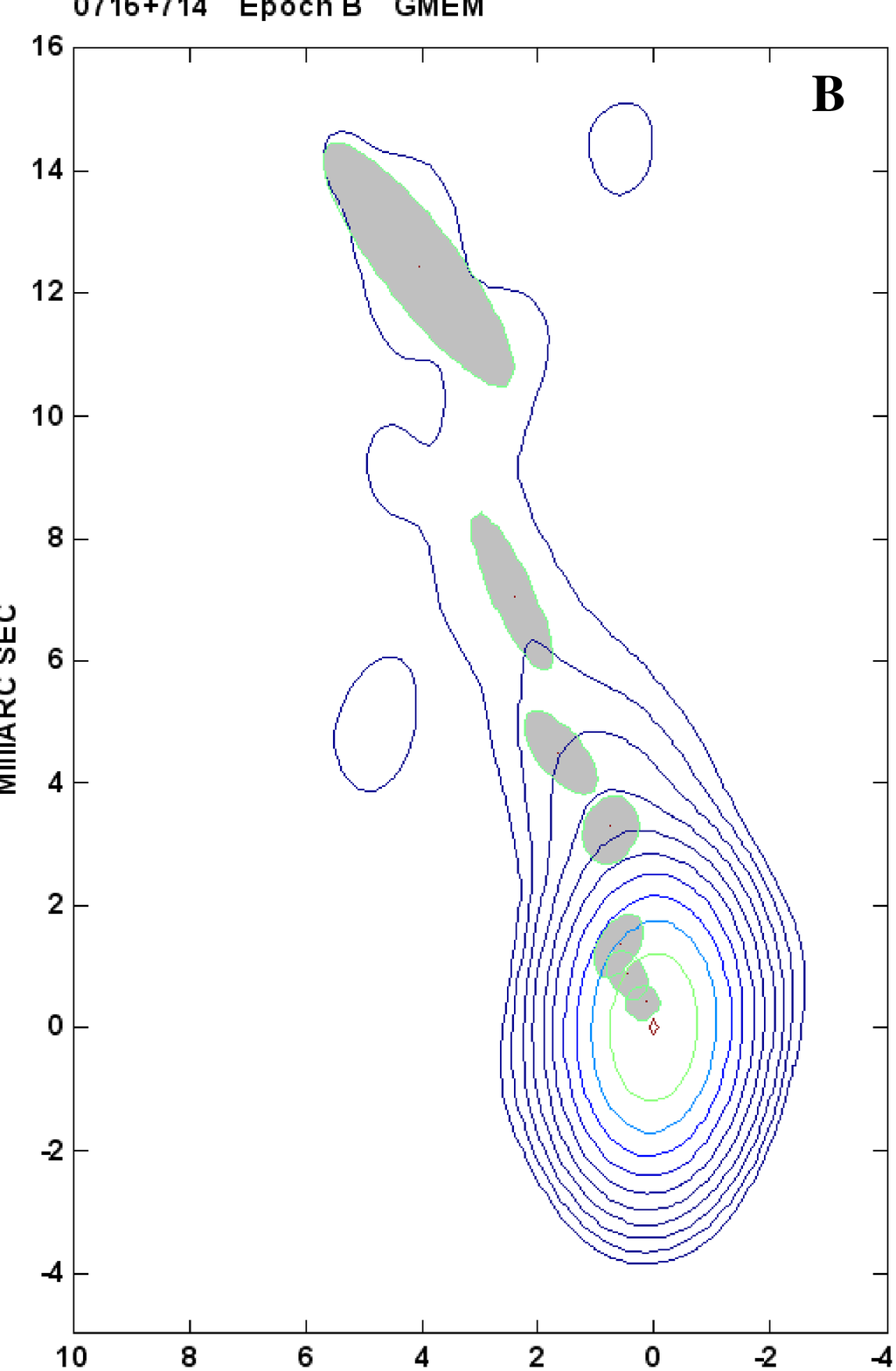}
\includegraphics*[angle=0,width=0.31485\textwidth]{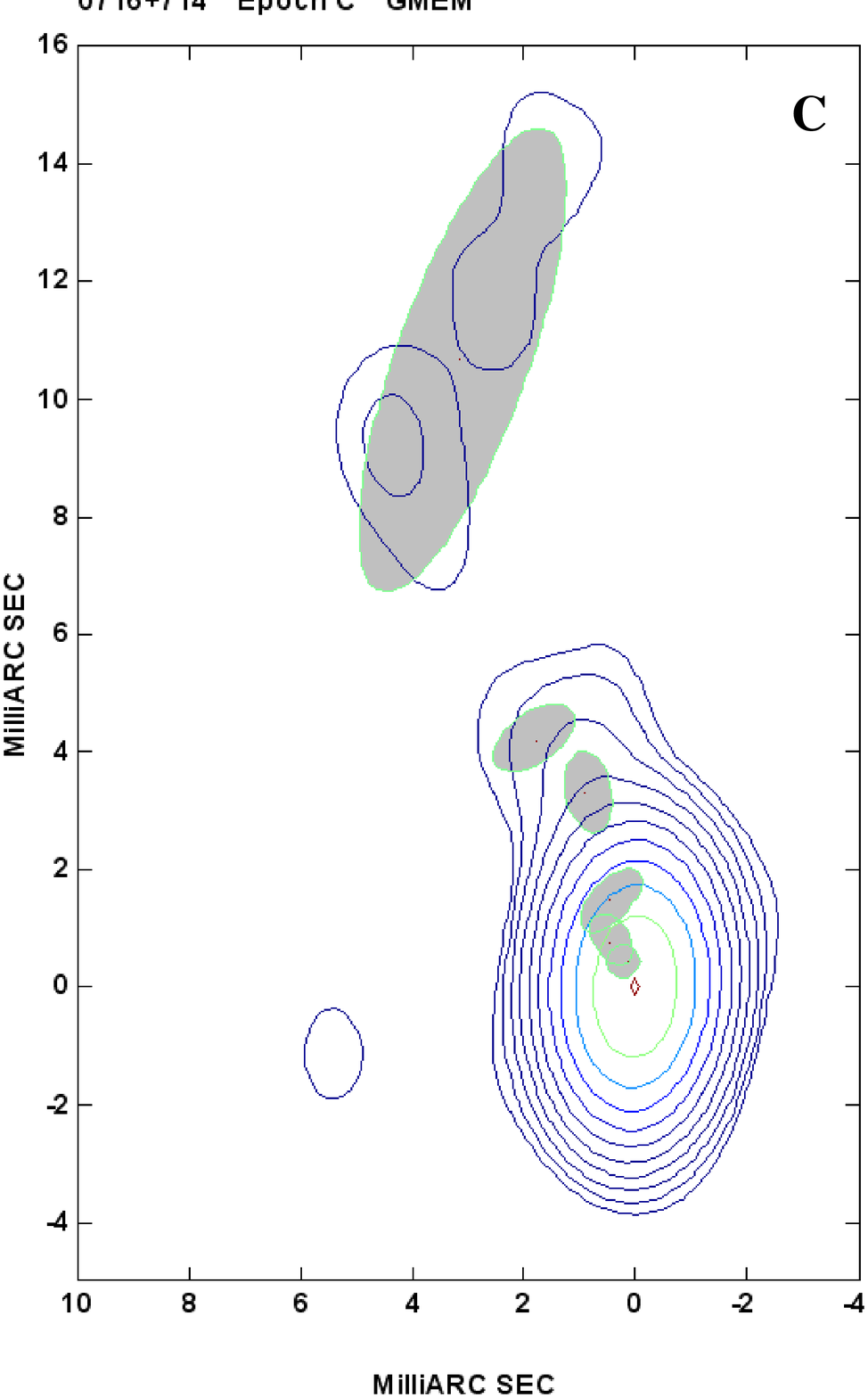}
\includegraphics*[angle=0,width=0.30\textwidth]{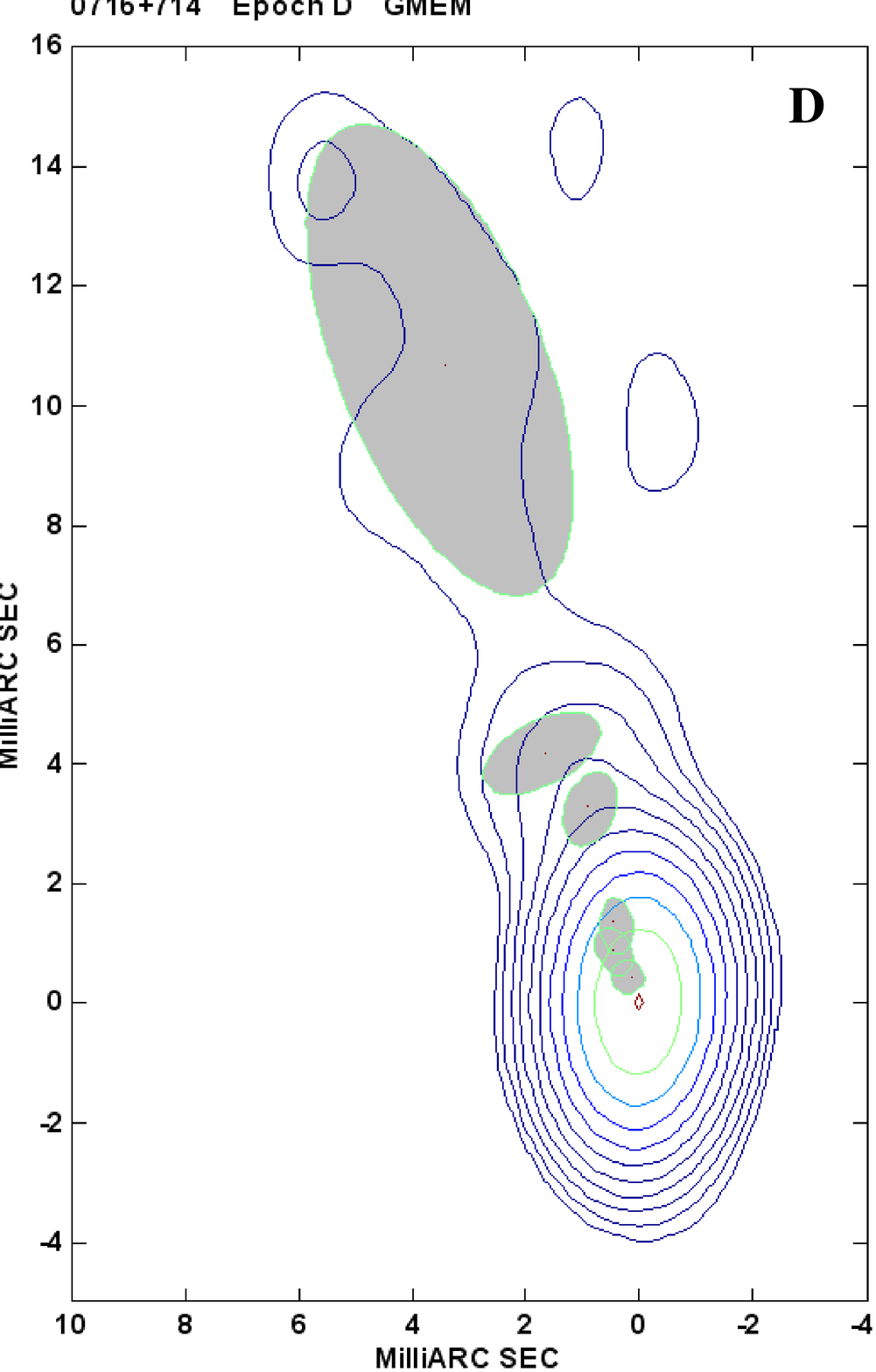}
\includegraphics*[angle=0,width=0.30\textwidth]{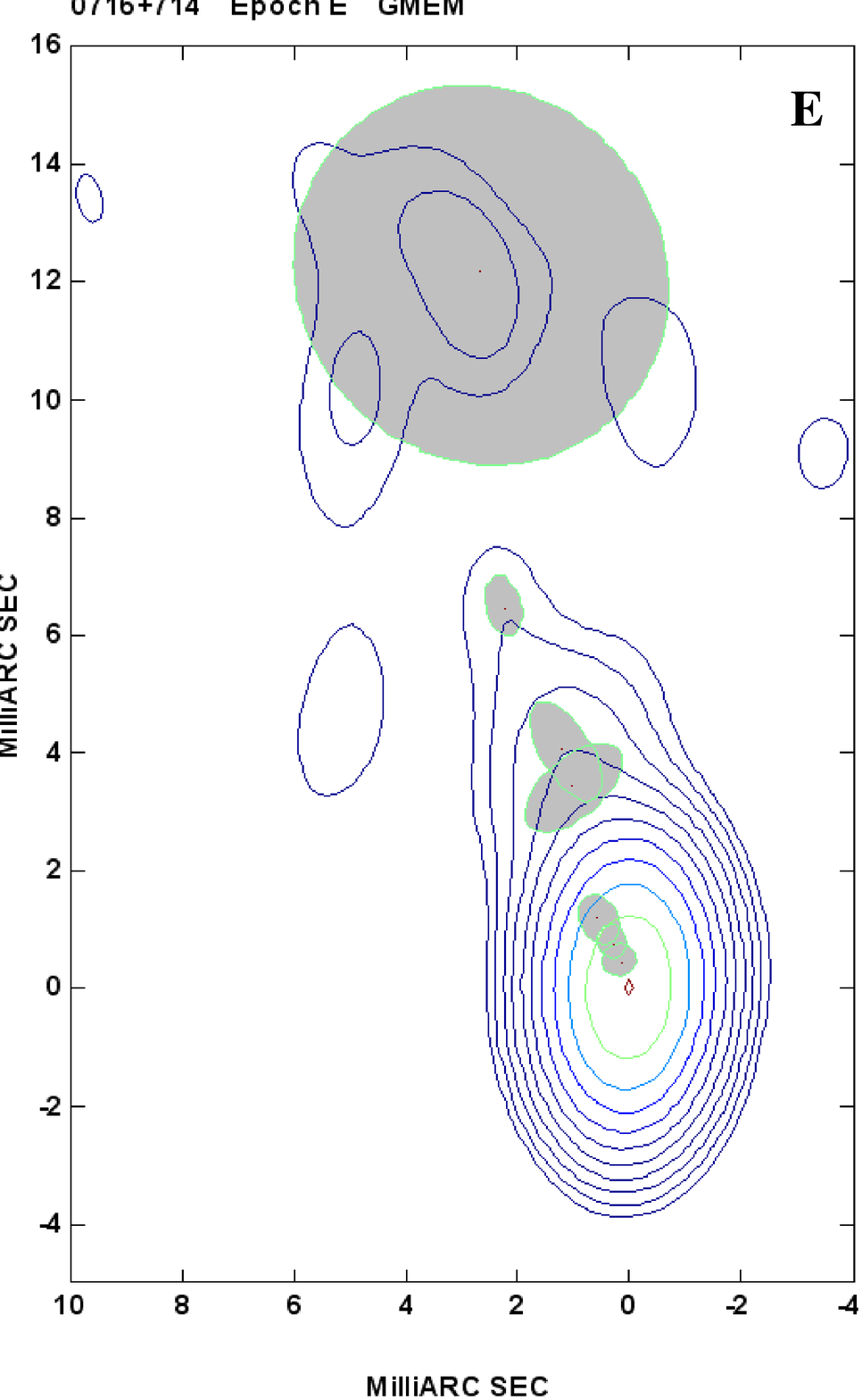}
    \caption{GMEM solutions for each epoch, convolved with the averaged CLEAN beam: 1.5$\times$2.5 mas, PA $= -4.5^{\circ}$. Gray ellipses are the Gaussian components superimposed on the image. The Gaussian model-fitting was performed in the image plane for the original GMEM solutions (Fig.~\ref{map:GMEMmaps0}) convolved with the beam of 0.5~$\times$~0.5~mas (Fig.~\ref{map:GMEMmaps05}). Axes are right ascension vs declination in milliarcseconds, measured with regard to the core.}
    \label{map:GMEMmaps}
\end{figure*}
After the {\em a priori} calibration, the raw visibility data were averaged in frequency (averaged IFs) in \textit{AIPS}, then exported to \textit{Difmap}, where they were rebinned, phase self-calibrated with a point source model with a 0.5 minute solution interval, and finally averaged in time (over a period of four minutes). After that the data was exported to \textit{VLBImager}, where the automatic GMEM deconvolution \-- self-calibration loop iterations were performed until convergence. The size for the maps is chosen $256\times 256$ pixels with a pixel size of 0.15 mas. The GMEM solution convolved with a 
clean beam of $\sim 2.4\times 1.5$ mas, $PA = -4.5^{\circ}$ is presented in Fig.~\ref{map:GMEMmaps}, and parameters for those maps are listed in Table~\ref{tb:GMEM_map_parameters}.  

\begin{table}[ht]
\caption[]{Map parameters for the GMEM. The two tables correspond to two different sizes of the beam, with which the original GMEM solutions were convolved.}\label{tb:GMEM_map_parameters}
\centering
\begin{tabular}{ccccc}\hline
Epoch & $S_{tot}$ &-   Beam                & $S_{peak}$ & $3\sigma$ \\
      & [Jy]      & mas$\times$ mas,[$^o$]& [Jy/beam]  & [mJy/beam] \\
\hline
2004.11 & 1.48 &$2.4\times$1.5,-4.5 & 1.42 & 0.97 \\
2004.22 & 1.08 &$2.4\times$1.5,-4.5 & 1.03 & 0.96 \\
2004.34 & 1.08 &$2.4\times$1.5,-4.5 & 1.00 & 0.74 \\
2004.46 & 1.15 &$2.4\times$1.5,-4.5 & 1.09 & 0.99 \\
2004.58 & 0.90 &$2.4\times$1.5,-4.5 & 0.82 & 0.59\\
2004.66 & 0.86 &$2.4\times$1.5,-4.5 & 0.78 & 0.67\\
\hline
2004.11 & 1.48 &$0.5\times$0.5,0 & 1.08 & 0.33 \\
2004.22 & 1.08 &$0.5\times$0.5,0 & 0.73 & 0.29 \\
2004.34 & 1.08 &$0.5\times$0.5,0 & 0.70 & 0.24 \\
2004.46 & 1.15 &$0.5\times$0.5,0 & 0.74 & 0.30 \\
2004.58 & 0.90 &$0.5\times$0.5,0 & 0.53 & 0.19\\
2004.66 & 0.86 &$0.5\times$0.5,0 & 0.48 & 0.19\\
\hline
\hline
\end{tabular}
\end{table}

\begin{figure*}
    \centering
    \includegraphics[angle=0,width=0.33\textwidth]{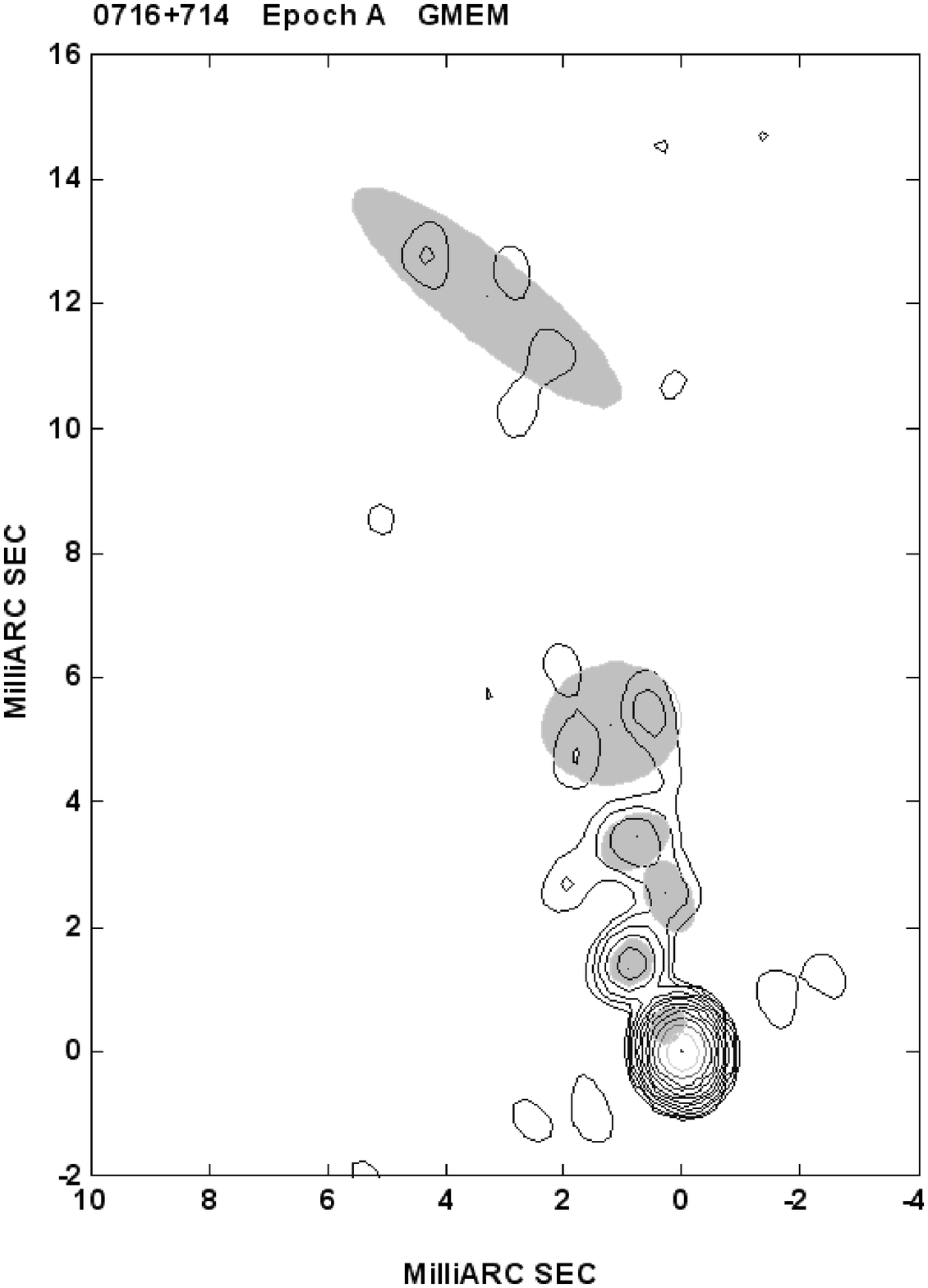} \includegraphics[angle=0,width=0.33\textwidth]{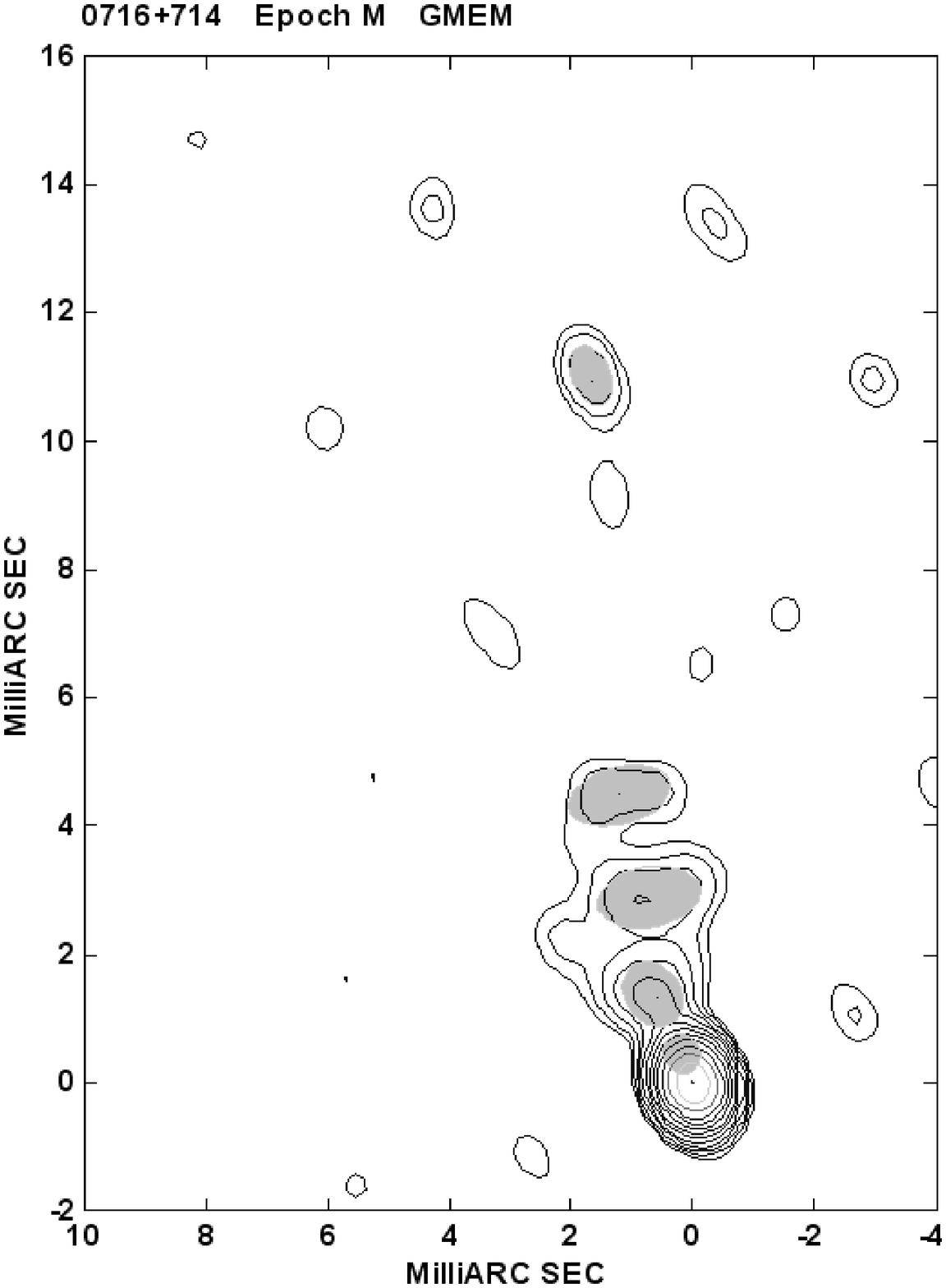} \includegraphics[angle=0,width=0.33\textwidth]{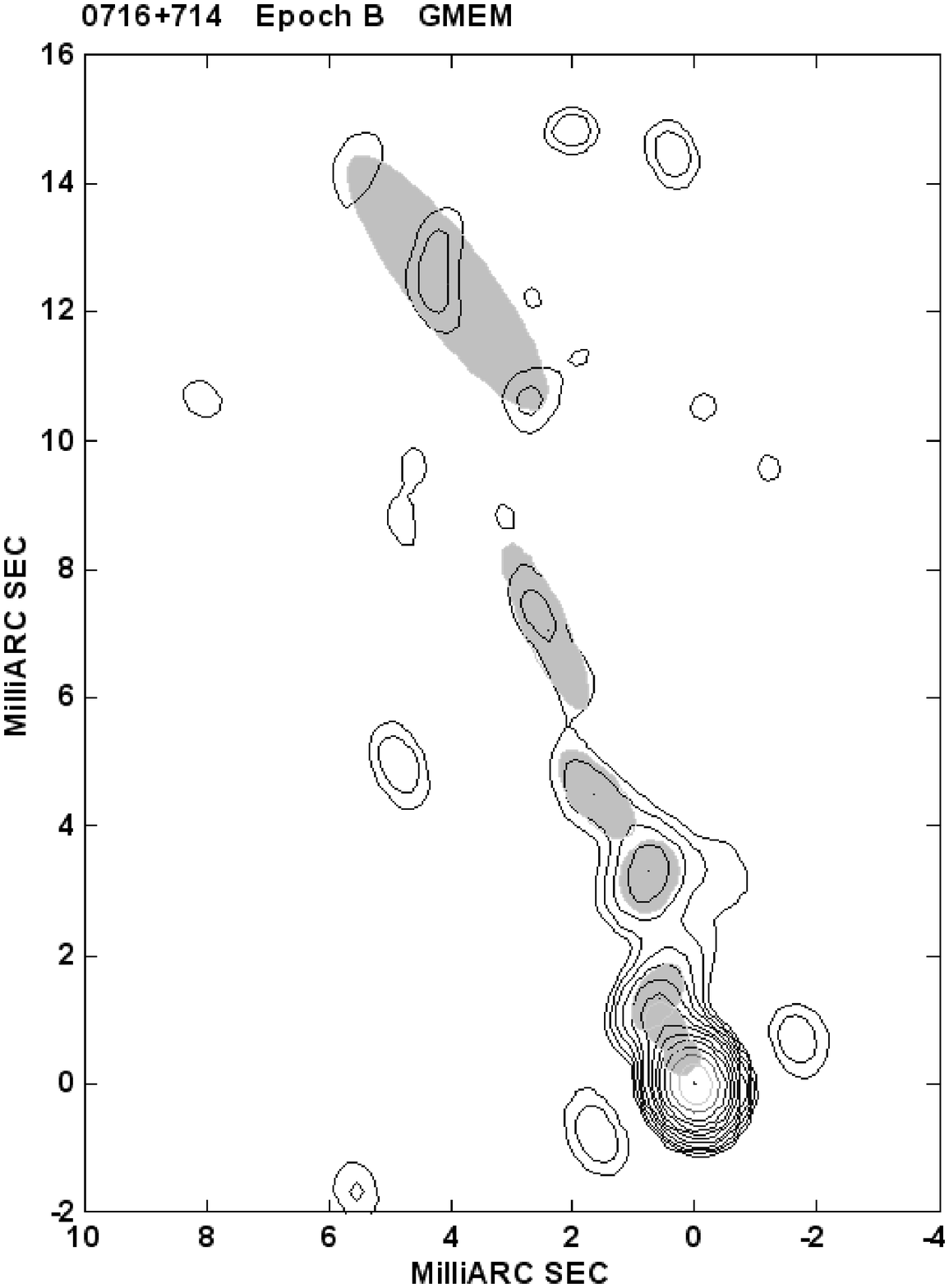}
\includegraphics[angle=0,width=0.33\textwidth]{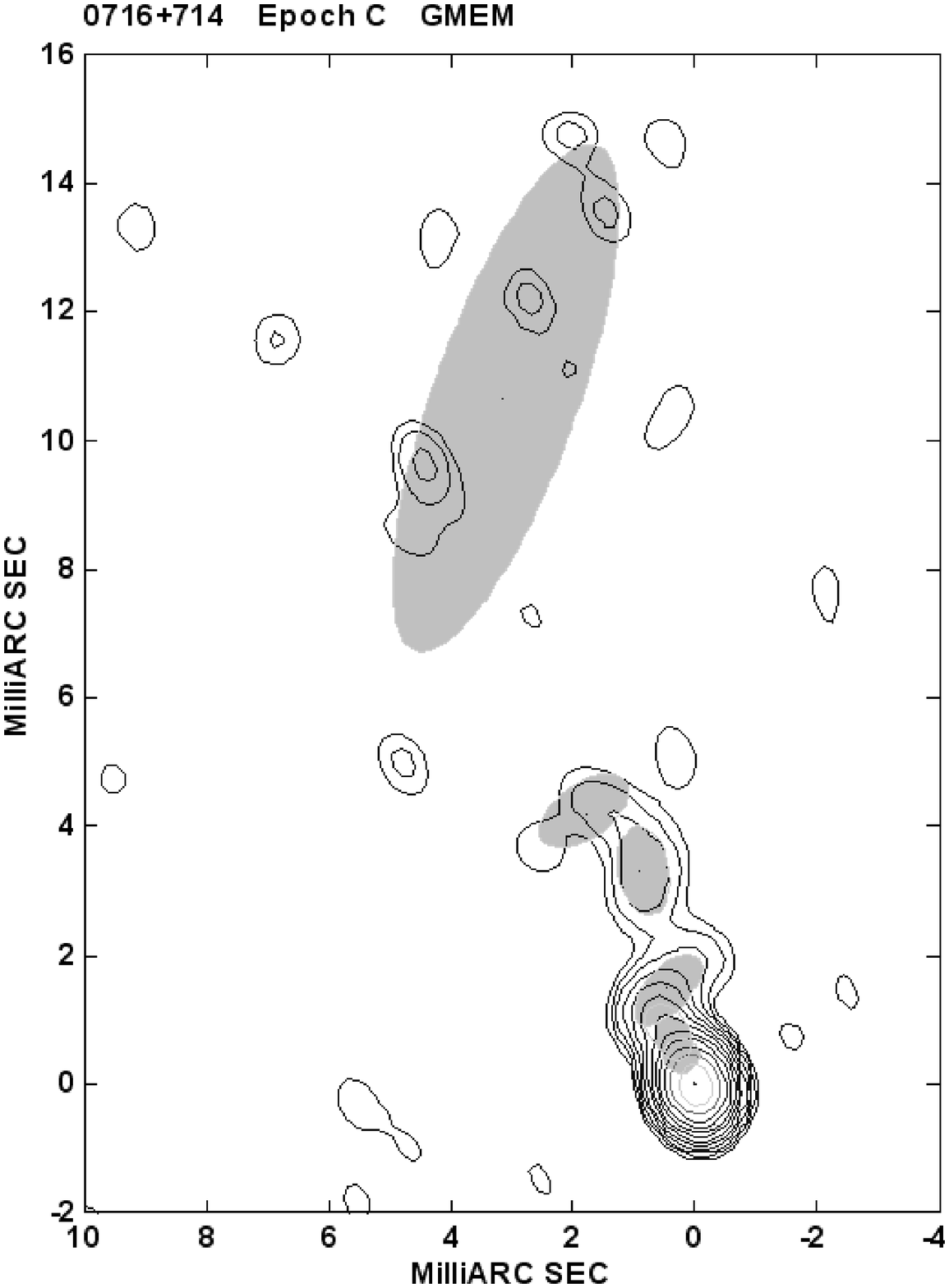} \includegraphics[angle=0,width=0.33\textwidth]{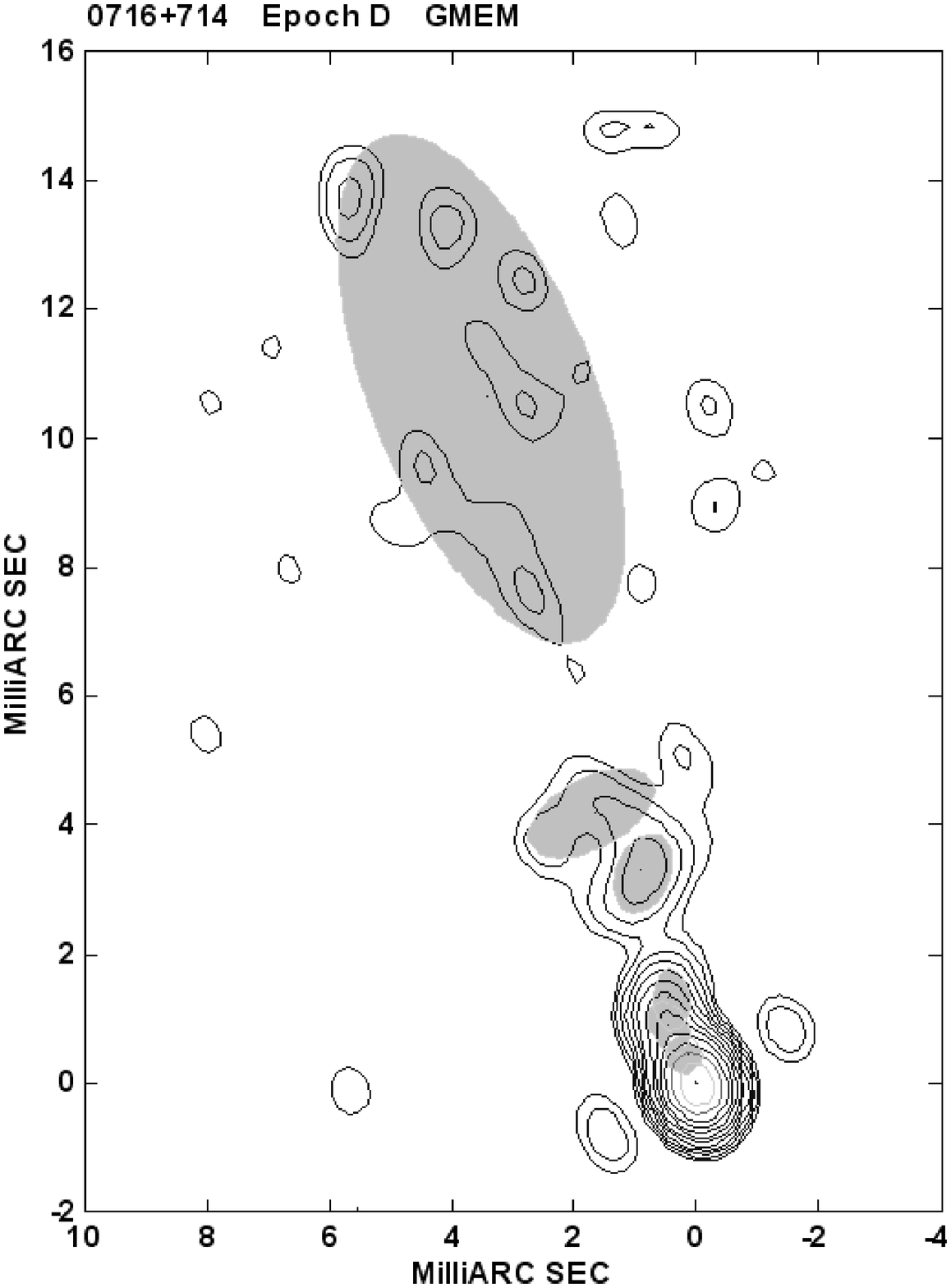} \includegraphics[angle=0,width=0.33\textwidth]{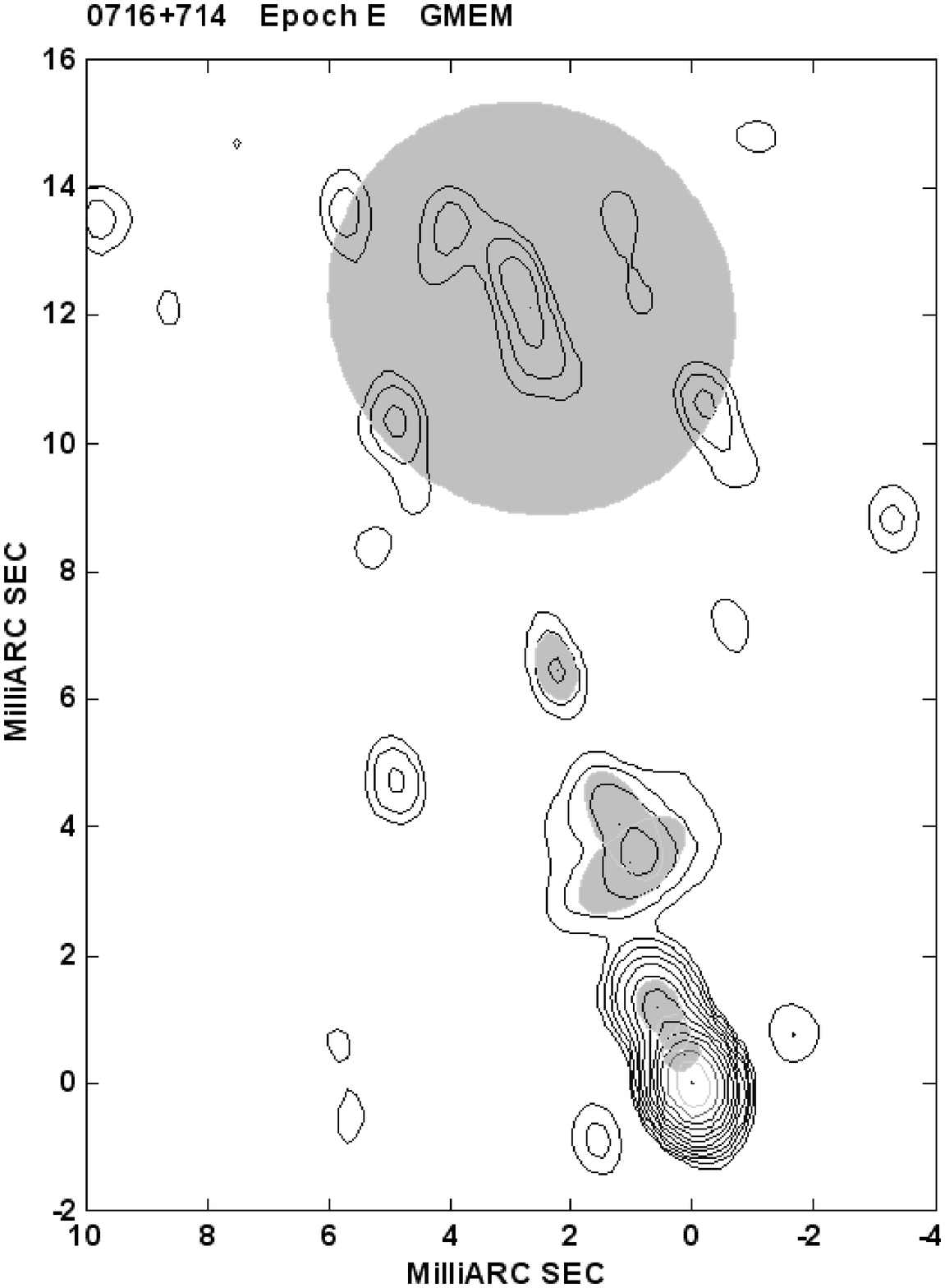}
    \caption{GMEM solutions, convolved with the beam of size 0.5x0.5 mas.}
    \label{map:GMEMmaps05}
\end{figure*}

In the appendix we present two additional sets of maps to demonstrate the capabilities of this deconvolution method: the original GMEM solutions (Fig.~\ref{map:GMEMmaps0}) and the GMEM solutions convolved with a circular clean beam of the size $0.5\times 0.5$ mas, which were used for the modeling of the source structure (Fig.~\ref{map:GMEMmaps05}).
%In order to study jet
%kinematics with required accuracy, maps with higher resolution
%were obtained by convolving MEM-solution with smaller beam of the size
%$0.5\times\ 0.5$ mas (Fig.~\ref{GMEMmaps05}).
The model-fitting of the maps was made in the image plane. The source structure was modeled using Gaussian elliptical components. Each of the components was described by the following six parameters: $r$, distance from the core (the core was determined like in the conventional method and was assumed to be stationary); $\phi$, position angle measured  counter-clockwise between the ordinate axis and the radius-vector; $a_\mathrm{maj}$ and $a_\mathrm{min}$, sizes of the major and minor
axes of a Gaussian ellipsoid at half maximum amplitude and $\theta$, the angle describing the
orientation of the component ellipse, an angle between the ordinate axis and the major axis of the ellipse, measured counter-clockwise. Cross-identification of the components was based
on the assumption that the core separation, the position
angle and the flux of the component change smoothly from epoch to epoch. In order to obtain
a model of the inner jet (r$<$1.5 mas) with higher accuracy,
we interpolated the brightness distribution to a finer grid: the pixel size in thit case was 0.0375
mas, which was four times smaller than that for the large-scale jet.
The errors of the component core separation determination were estimated by the formula \citep{fomalont1999}

\begin{eqnarray}
&& \Delta r=\frac{\sigma\Theta}{2S_\mathrm{peak}}~,\nonumber
\end{eqnarray}
where $\sigma$ is the rms after the subtraction of the model from the
source map, $S_\mathrm{peak}$ is the peak flux of the component, $\Theta$ is the
size of the component at the amplitude half maximum (in our case we assume $\Theta=a_\mathrm{maj}$). 
We point out that that this formula likely underestimates position errors, which is clear from the comparison with the {\textit Difwrap} errors (See Fig.~\ref{all_components}).
%To determine errors in proper motion, a Monte-Carlo simulation was used with a
%population size of 3000 random component positions normally distributed within calculated error limits.
%error limits given in the Table~\ref{gmem_modelling} were generated.

\section{Results}\label{Results}
Our aim was to study the structure and kinematics of the large-scale
(from 1 to 12~mas) jet, and compare it with the results for the inner ($<1$~mas) jet, which were published in  \citet{rastorgueva2008}.

We applied a conventional method and a GMEM to the same data set to compare their kinematic results. For each of them, we calculated the core separation of the components as a
function of time, the proper motion $\mu$, and the apparent speeds $\beta_{app}$
according to the formula

\begin{eqnarray}
&& \beta_{app}=\frac{\mu d_L}{c(1+z)}~,\nonumber
\end{eqnarray}
where the luminosity distance $d_L$ is calculated with the analytical formula derived by
 \citet{pen1999} for a flat cosmology. The same analytical expression was used by
\citet{bach2005} for their kinematic analysis of this source, therefore our results
could be compared directly without conversion. For this cosmological model,
a proper motion of 1 mas/yr is equivalent to the apparent speed of 19.26\textit{c}. The
flat Universe model with the Hubble constant of $H_0=71 $km/s/Mpc and with $\Omega_m=0.3$
and $\Omega_{\Lambda}=0.7$ was used. We are aware that \citet{wick2010} recently obtained a new and more precise analytical expression for the luminosity distance, but for
z=0.3 the difference with \citet{pen1999} is negligible (1 mas/yr corresponds to
the apparent speed of 19.22\textit{c}).

%\begin{figure}[t]
%    \centering
%    \includegraphics[angle=0,width=6cm,height=6cm]{pics/slow_fit_CLEAN_FINAL.ps}
%    \caption{Motion of the jet components, obtained using the conventional method.}
%        \label{fit:CLEANfit}
%\end{figure}

\begin{table}
\caption[]{Kinematic scenario obtained with the conventional method. The table contains the component name (enumeration starts from the vicinity of the core), the average distance from the core in mas, the component proper motion in mas/yr and apparent speed in the units of speed of light.}
\label{app_speeds_clean} \centering
\begin{tabular}{cccc}
\hline\hline
Comp. & Core dist.& $\mu$  & $\beta_{app}$   \\
name  & [mas] & [mas/yr] &     [$c$]    \\
\hline
C1-2 & 0.64$\pm$0.14 & 0.55$\pm$0.11 & 10.6$\pm$2.2  \\ % a5
C3 & 1.9$\pm$0.4 & 0.62$\pm$0.41 & 11.8$\pm$8.0 \\ % a4
C4 & 3.2$\pm$0.4 & 0.44$\pm$1.4 & 8.6$\pm$26.5   \\ % a3
C5 & 5.7$\pm$0.9 & -1.37$\pm$2.7 & -26.4$\pm$52.2 \\ % a12
C6 & 11.6$\pm$0.6 & -0.27$\pm$1.2 & 5.2$\pm$23.6  \\ % a0
\hline
\end{tabular}
\end{table}

%\begin{figure}[h!]
%    \centering
%    \includegraphics[angle=0,width=6cm,height=6cm]{pics/fit_gmem_err.ps}
%    \caption{Motion of the jet components, obtained using GMEM.}
%    \label{GMEMfit}
%\end{figure}

\begin{table}
\caption[]{Kinematic scenario obtained with the GMEM. The table contains the component name (enumeration starts from the vicinity of the core), the average distance from the core in mas, the component proper motion in mas/yr and apparent speed in the units of speed of light.} \label{gmem_speeds}
\centering
\begin{tabular}{cccc}\hline
Comp. & Core dist.& $\mu$  & $\beta_{app}$   \\
name  & [mas] & [mas/yr] &     [$c$]    \\
\hline
A1 & 0.49$\pm$0.02 & 0.12$\pm$0.05& 2.4 $\pm$0.9\\
A2 & 0.93$\pm$0.06 &-0.24$\pm$0.40&-4.6 $\pm$7.1\\
A3 & 1.48$\pm$0.13 &-0.64$\pm$0.03&-12.4$\pm$0.6\\
A4 & 3.3$\pm$0.2 & 0.13$\pm$0.32& 2.4 $\pm$6.2\\
A5 & 4.7$\pm$0.4 &-1.37$\pm$0.37&-26$\pm$7\\
$A6^{*}$ & 11.9$\pm$0.8 &-0.45$\pm$1.13&-9$\pm$20\\
\hline
\end{tabular}
\begin{list}{}{}
\item[$^*$] fit performed without taking into account errors of the component position estimate
\end{list}
\end{table}

\begin{figure*}[h]
    \centering
    \includegraphics[angle=0,width=0.33\textwidth]{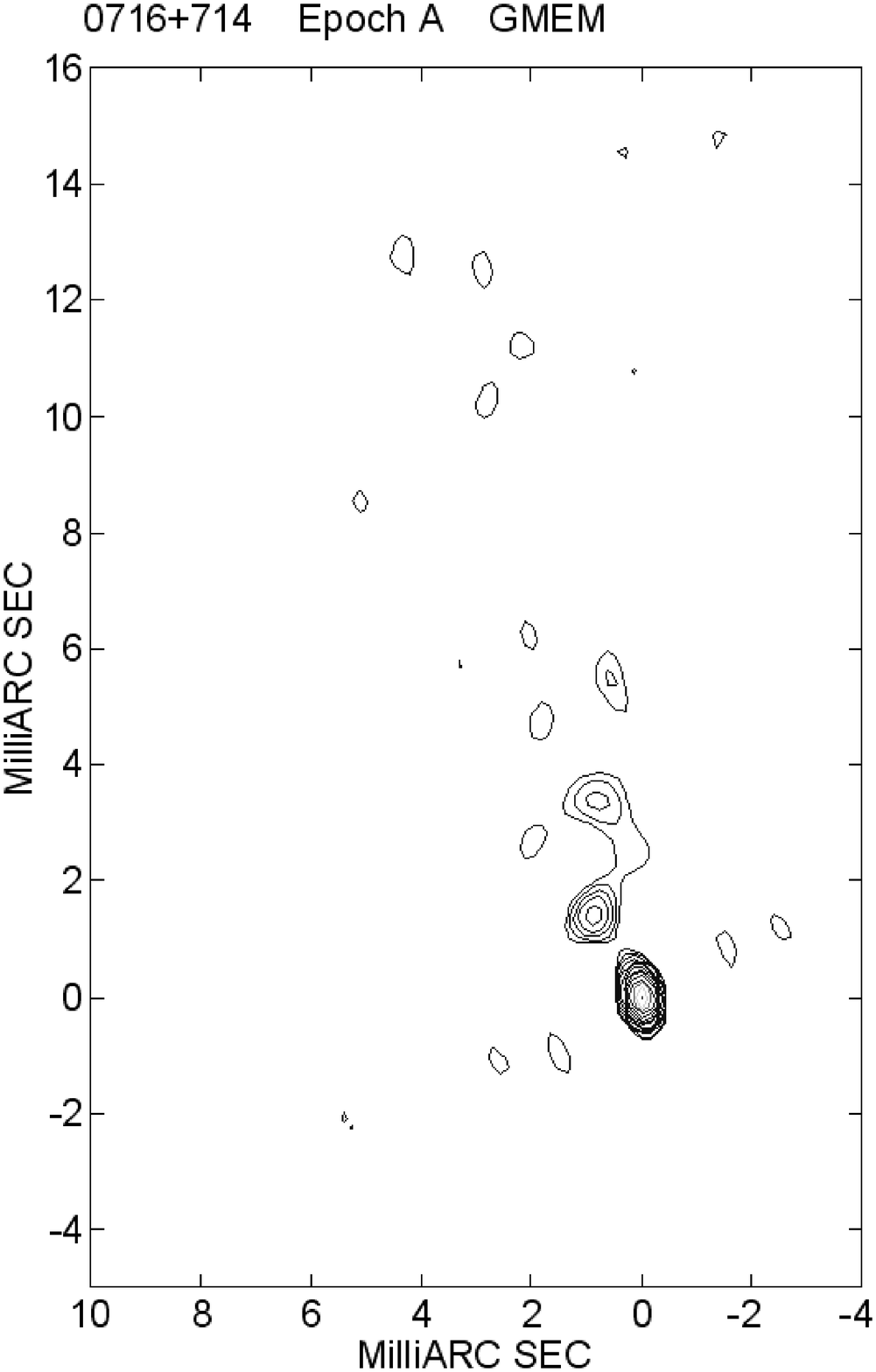} \includegraphics[angle=0,width=0.33\textwidth]{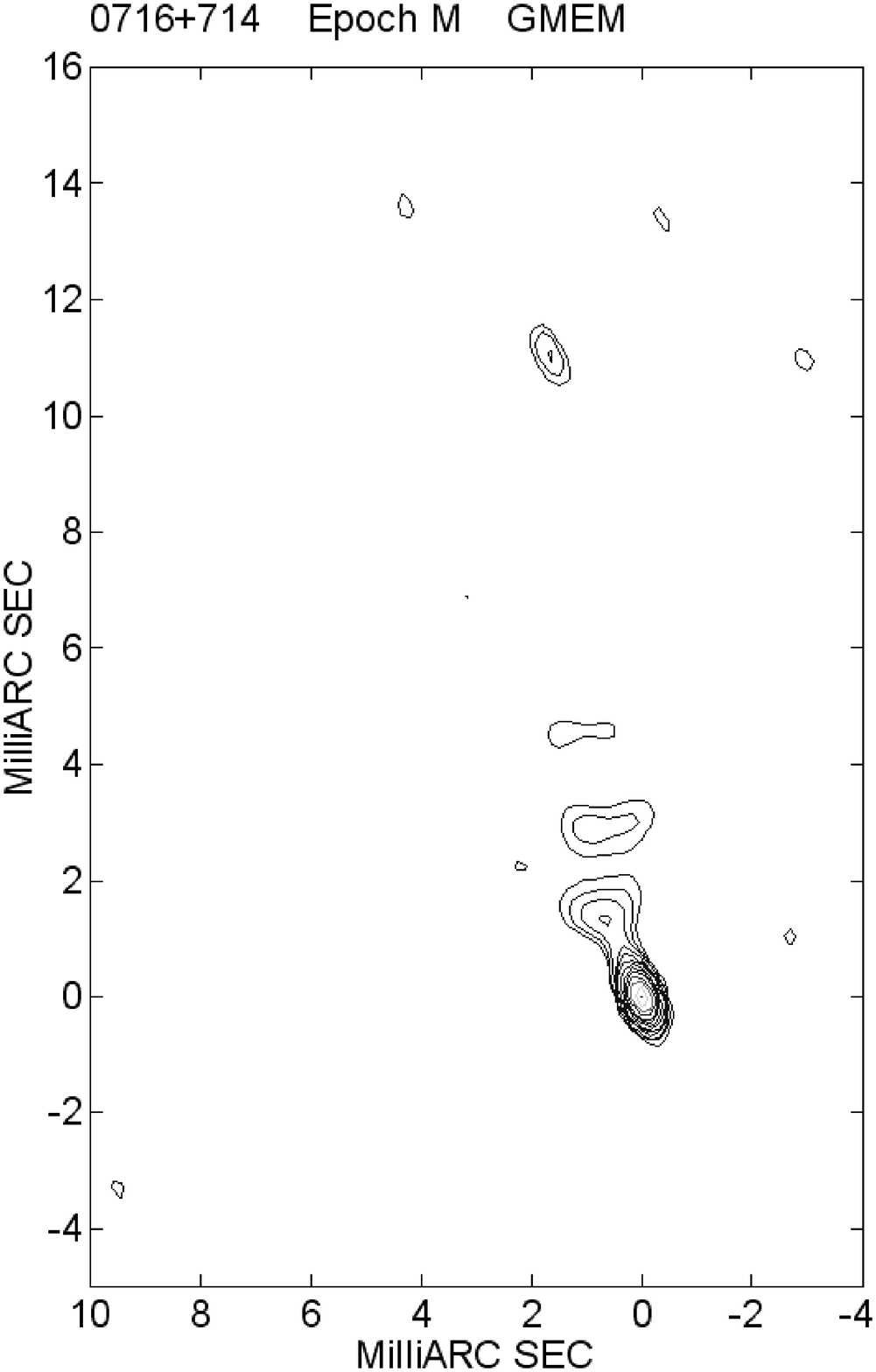} \includegraphics[angle=0,width=0.33\textwidth]{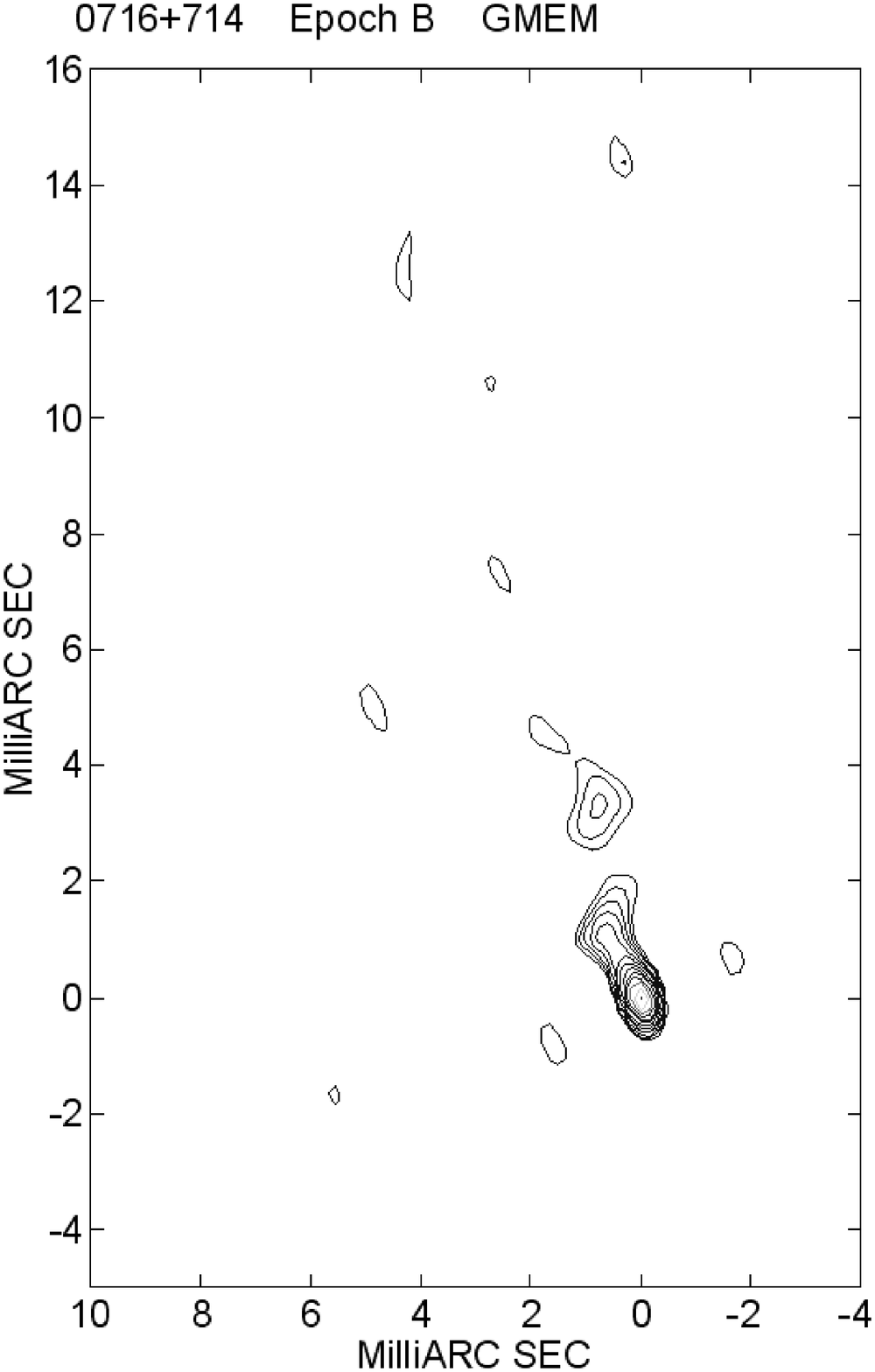}
\includegraphics[angle=0,width=0.33\textwidth]{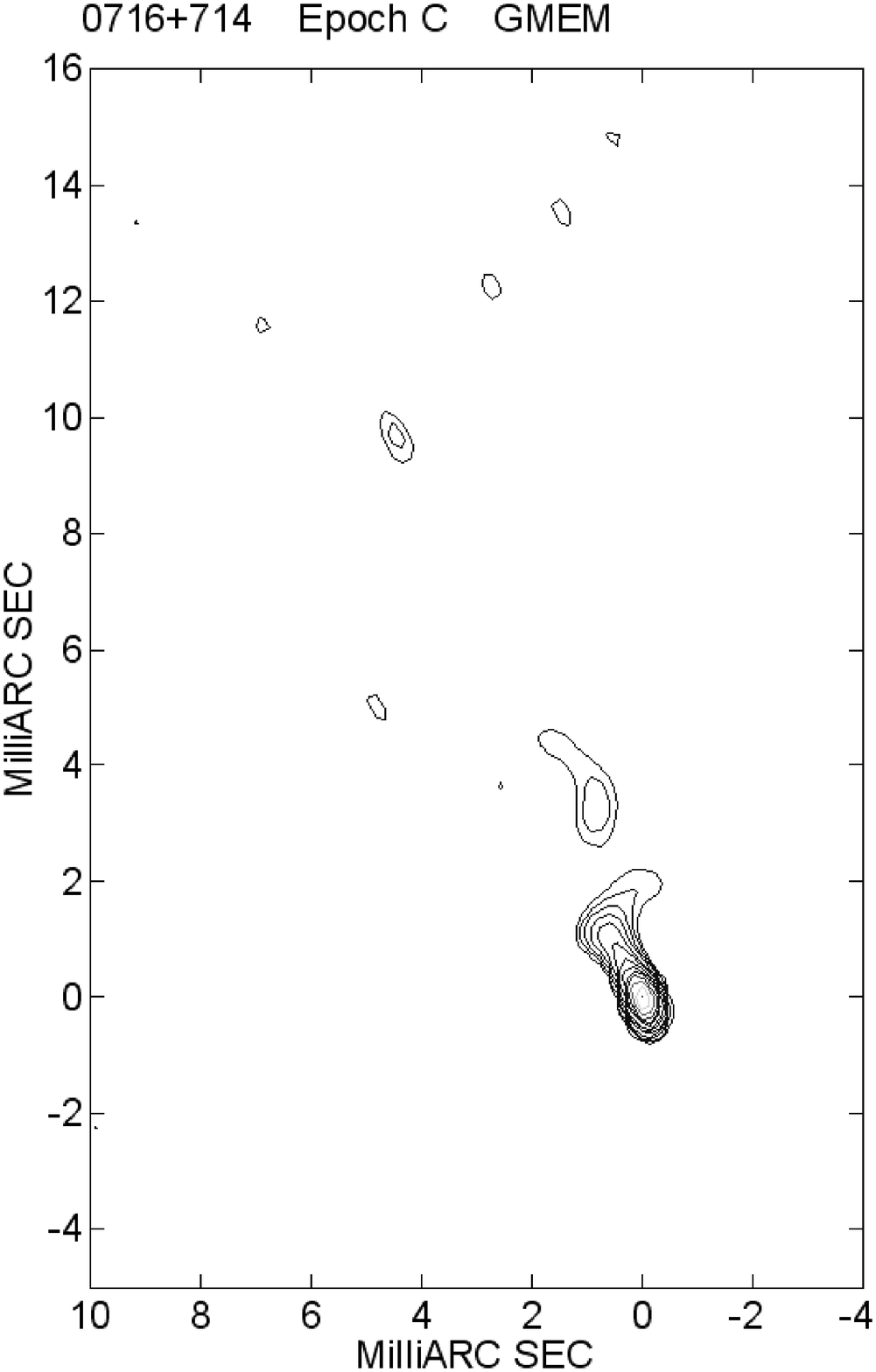} \includegraphics[angle=0,width=0.33\textwidth]{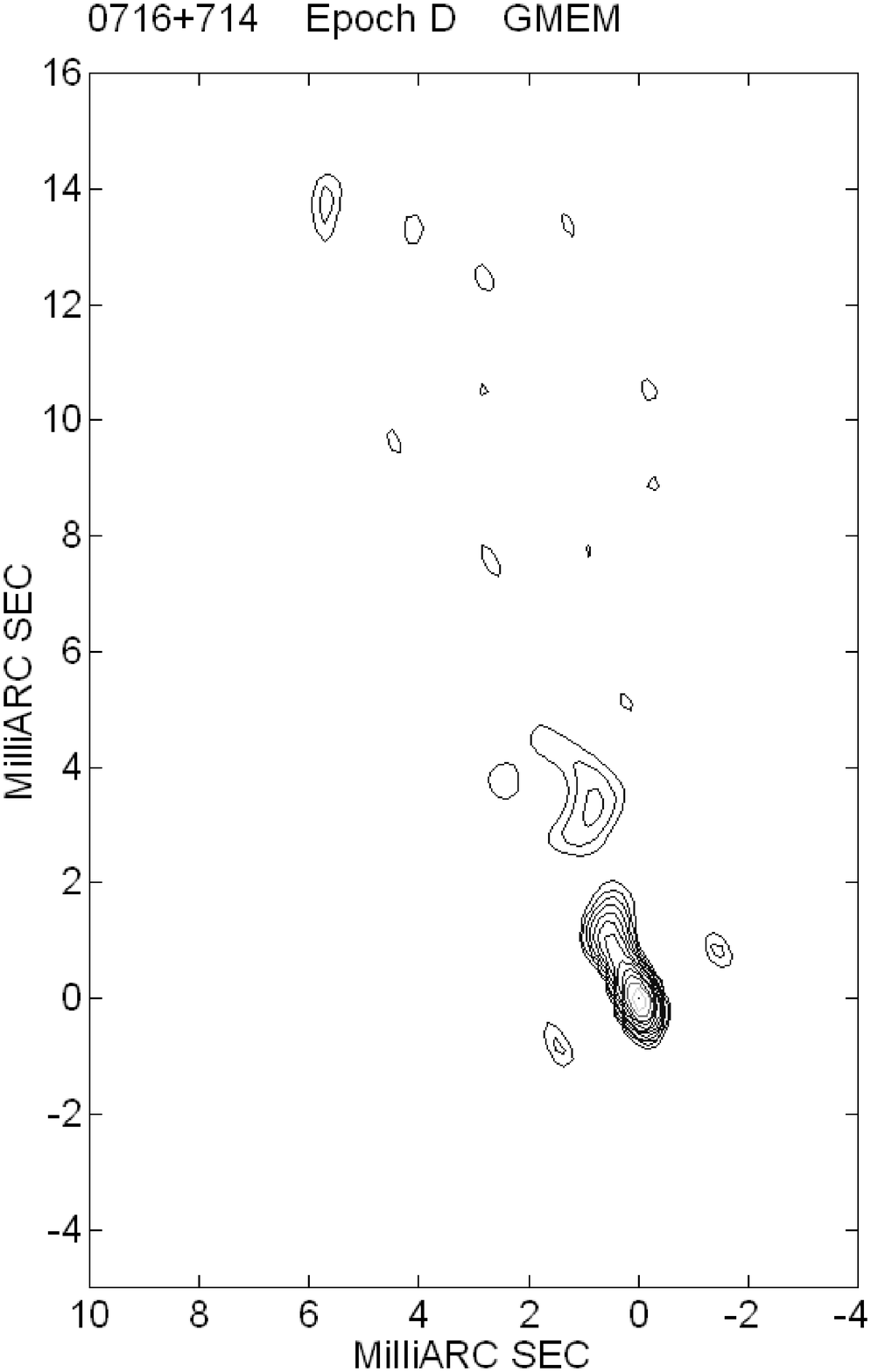} \includegraphics[angle=0,width=0.33\textwidth]{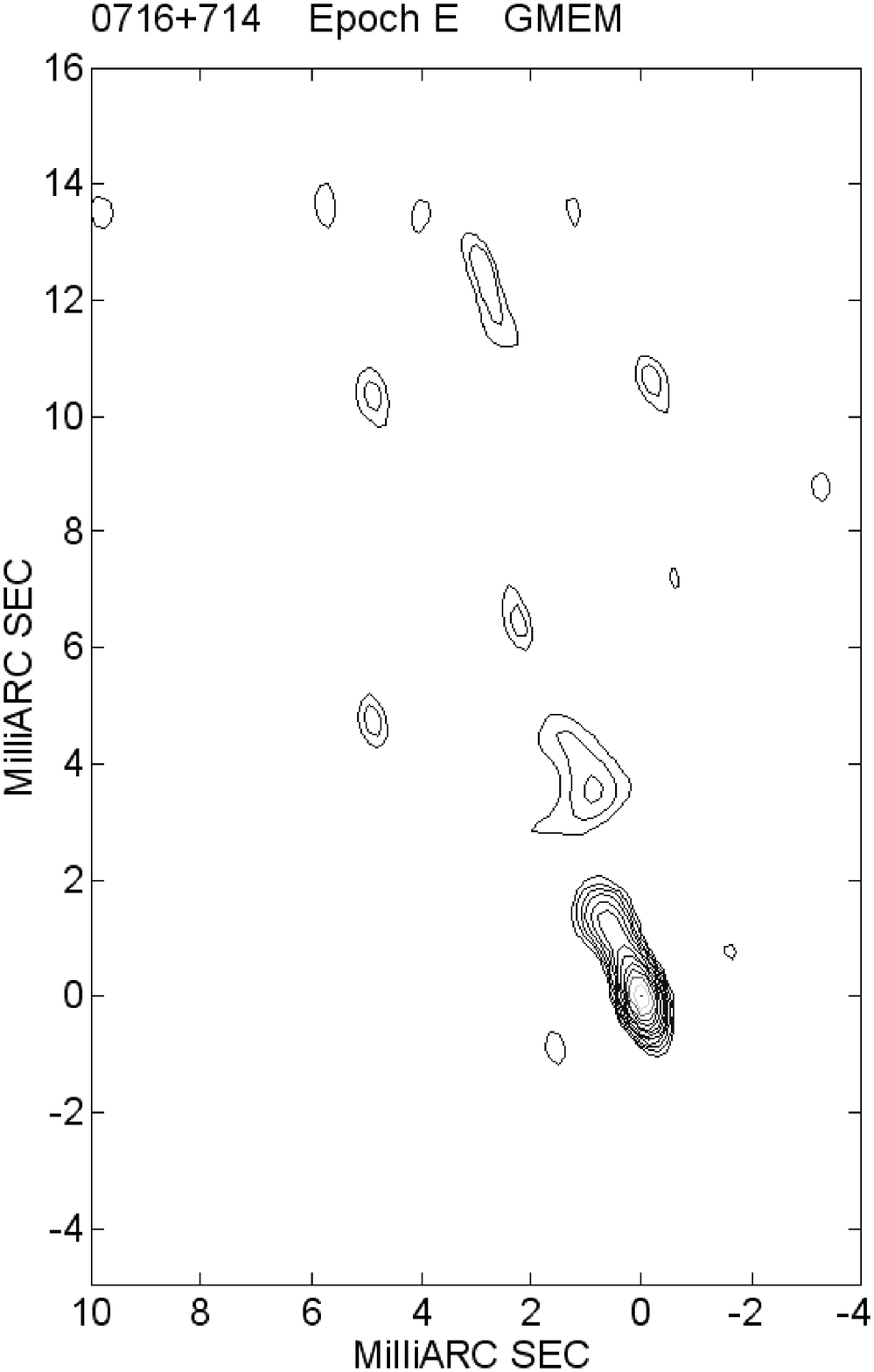}
    \caption{Original GMEM solutions.}
    \label{map:GMEMmaps0}
\end{figure*}

The results of the Gaussian model-fitting and component
identification are presented in Table~\ref{tb:CLEAN_models} (conventional method) and in the Table~\ref{tb:GMEM_models} (GMEM). The original
 GMEM-solutions are shown in Fig.~\ref{map:GMEMmaps0}. %and the same image convolved

\subsection{Component cross-identification between different methods.} 
\subsubsection{Positions and apparent speeds of components}
\begin{figure}[h!]
    \centering
    \includegraphics[angle=0,width=8cm]{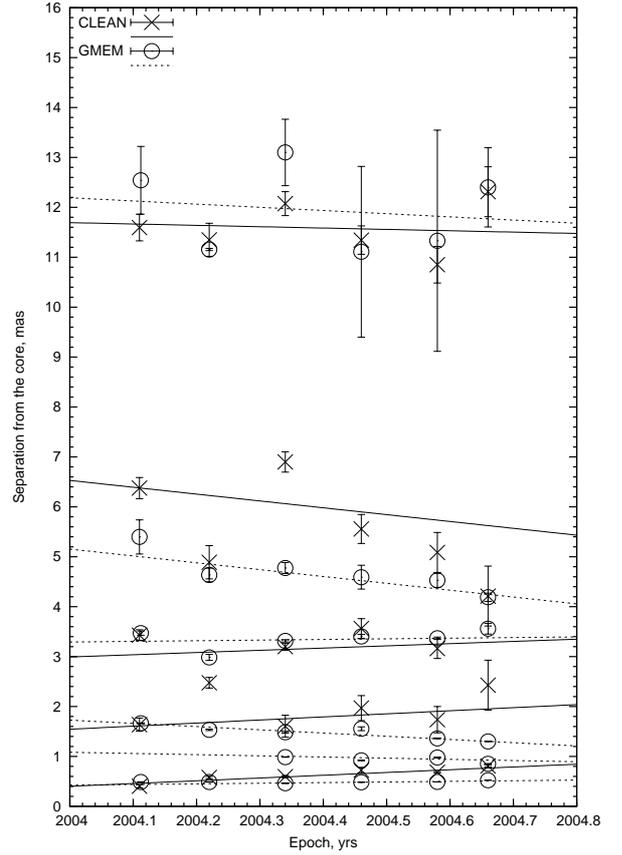}
    \caption{All components obtained with the conventional and GMEM methods are shown at the same plot.}
    \label{all_components}
\end{figure}

Based on core separation and fluxes, we identified four nearly stationary components in the large-scale jet of 0716+714: one at a separation of about $\sim 1.5$~mas (C3 \-- A3), one at $\sim $3.0~mas (C4 \-- A4), one at $\sim $4.6~mas (C5 \-- A5) and one at $\sim $12~mas (C6 \-- A6). Their proper motions and apparent speeds can be found in Tables~\ref{app_speeds_clean} and~\ref{gmem_speeds}. A comparison of their positions, obtained with different methods, and linear fits are shown in Fig.~\ref{all_components}. The components C1-C6, found by the conventional method, have a larger scatter around the mean position and larger positional errors than those of the GMEM components (compare Tables~\ref{app_speeds_clean} and~\ref{gmem_speeds}), which complicated the component cross-identification and made a choice of the kinematic scenario ambiguous. However, the GMEM method confirmed the stationary scenario.

However, the component C1-2, determined by the conventional method, and the component A1 found by the GMEM in the inner jet have a clear outward motion: $\sim 10c$ and $\sim 2c$, respectively (Tables~\ref{app_speeds_clean} and~\ref{gmem_speeds}). The difference in apparent speed could be explained by the fact that the GMEM determined an extra stationary component at the separation of about 1 mas from the core (A2), while the conventional method did not. On the other hand, A2 is only found at the last four epochs, and is separated from the component A1 by only $\sim 0.5$~mas, and the component C1-2, determined by the conventional method, is located between them (see Fig.~\ref{all_components}). Therefore, we believe that the component C1-2 is resolved by GMEM in the four last epochs, making up two components A1 and A2, and this effect causes the difference in the apparent speed.

Note that the kinematics of the outer jet are consistent with the results of \citet{britzen2009}.  

\subsubsection{Core flux}
In this section we compare the ability of the two methods to restore the flux of the source. In the optically thin jet, the fluxes of the components found by the conventional method and by the GMEM agree within two sigma. At the same time, the performance of the two methods in the core region is significantly different: the conventional method places one very bright  component C1-2 (the flux density is about 0.1 Jy, so it is bright compared to other jet components) next to the core, while the GMEM method finds two components, A1 and A2, with the flux density comparable to the other components in the jet (on the order of 0.01 Jy). However, if one adds up the flux densities of the core and nearby components (C0+C1-2 and A0+A1+A2), the results of the two methods would agree within two sigma for all six epochs. We attribute this difference to the fact that the GMEM restores the bright point-like core accurately, whereas the conventional method, because of significantly lower resolution at 5 GHz, is likely to mix up emission of the core and  of the nearby components. 

\section{Discussion} \label{Discussion}
\subsection{Inner jet kinematics: comparison of the 22 and 5 GHz data}
\begin{figure}[h!]
    \centering
    \includegraphics[angle=0,width=8cm]{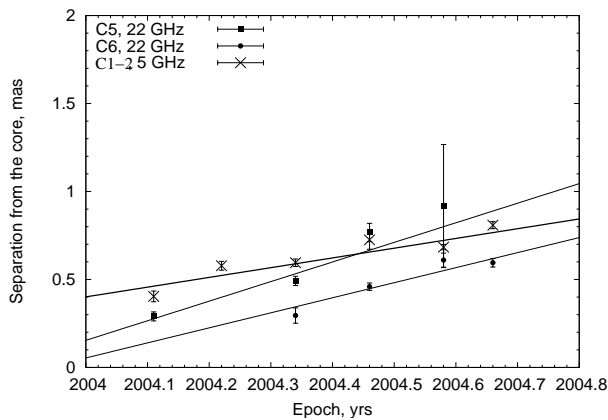}     
    \caption{22 GHz inner jet components C5 and C6 from \citet{rastorgueva2008} and the component C1-2 at 5 GHz coincide spatially and have similar motion.}
    \label{C5}
\end{figure}

The inner jet of 0716+714 is well resolved at higher ($>~5$~GHz) frequencies and reveals a clear outward motion, although the emission fades away at about 1 mas as the spectral peak moves to lower frequencies \citep[see][]{rastorgueva2008}. At 5 GHz, only one or two components are resolved in the inner jet region, depending on the method, and they also have a clear outward motion. Because the data at higher frequencies were reduced by the conventional method, we discussin in this section  only the component C1-2. Its apparent speed can be found in Table~\ref{app_speeds_clean}. In Fig.~\ref{C5} we show the motion of the component C1-2 at 5 GHz together with components C5 and C6 at 22 GHz. They are obviously located in the same area in the inner jet and have speeds of a similar values: the apparent speeds of the components C5(22 GHz) and C6(22 GHz) are 21.3$\pm$2.5 and 18.9$\pm$0.8 respectively, which agrees with the speed of C1-2(5 GHz) within $3 \sigma$. The difference in the apparent speeds of the inner jet components, detected at different frequencies, is likely caused by a resolution effect. 

\subsection{Jet bending and beaming}
\begin{figure}[h!]
    \centering
    \includegraphics[angle=0,width=8cm]{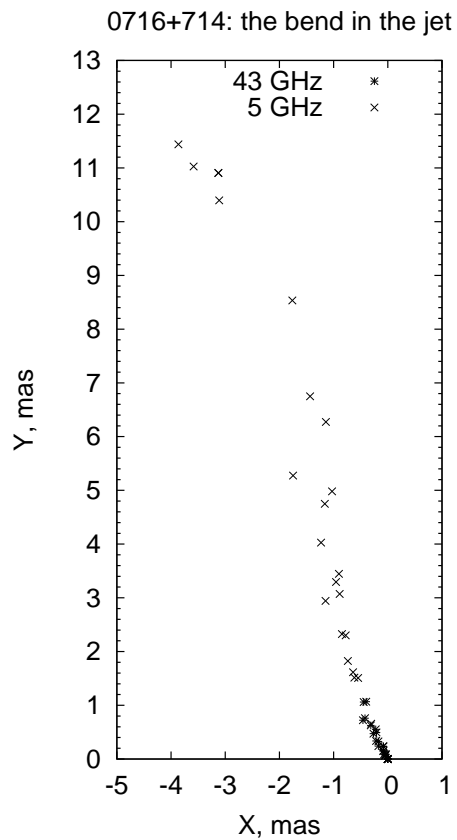}     
    \caption{Components of the inner and outer jet on the sky plane. The change of the orientation occurs at the distance of about 1~mas from the core. The components of the jet are well aligned, which is an indication of the collimated jet.}
    \label{jetbend}
\end{figure}

The jet components that were found beyond 1~mas from the core at both 5 and 22 GHz have a very high dispersion in the apparent speeds, and in most cases move slower than the inner ones (see Table~\ref{app_speeds_clean}). Therefore, we observe two kinematically different regions in the jet of 0716+714: a fast moving inner jet ($<1$ mas) and a slowly moving outer jet with nearly stationary components ($>1$ mas). The orientation of the inner and outer parts of the jet on the sky plane are also different: the jet changes its position angle from $\sim 25^{\circ}$ (inner region) to $\sim 15^{\circ}$ (outer region). The bend occurs at the distance of about 1~mas from the core, see Fig.~\ref{jetbend}. In addition to that, we found that the flux of the jet components in the inner jet is systematically higher than the flux of the outer jet components. Based on this we propose that those effects could be caused by the jet bending away from the line of sight at the apparent distance of 1~mas from the core.
We tried to find out whether it is possible to explain the observed effects by the jet bending, without additional assumptions of jet parameters. If the jet bends away from the line of sight, the beaming factor $\delta$ decreases, which causes a decrease in the apparent luminosity of the jet, because it is proportional to $\delta^{2+\alpha}$ for a smooth jet  \citep{cohen2007}, where $\alpha$ is a spectral index ($S \sim \nu^{\alpha}$). The apparent speed of the jet also changes with the change of the angle to the line of sight, for 0716+714 it decreases. We estimated how much the flux of the jet components change from the inner jet to the outer, and compared these results with the theoretical predictions for jet bending. We assumed the jet to be a smooth collimated flow with a constant plasma speed (and therefore constant Lorentz factor $\Gamma$). This assumption is indirectly supported by the fact that the jet components are well aligned (see Fig.~\ref{jetbend}), and the structure of the jet on VLBI maps is smooth and continuous. We used the results of the conventional method for this analysis. 

For each epoch we calculated the ratio of the average component flux at 5 GHz in the outer part of the jet to the average flux at 22 GHz of the inner ones: $r_{I} = <I_{o}>/<I_{i}>$. %The inner jet components were determined at 22 GHz,  and all but two outer jet components were derived from the 5 GHz models. 
Before calculating the flux ratio $r_{I}$ for each epoch, we corrected the spectral slope of the 5 GHz fluxes in order to compare them directly with the 22 GHz data. The 5-22 GHz spectral index $\alpha$ was estimated for the inner jet components C1-2 (5 GHz) and C5 and C6 (22 GHz) for each epoch separately. Individual values of the flux ratios were averaged, which gave  $<r_{I}> = 0.11\pm0.08$. Assuming that the flux change is caused by jet bending, we can calculate the ratio of the beaming factors of the jet after and before the bent using the flux ratio we have just obtained: $<r_{\delta}> = r_{I}^{2+\alpha} = 0.24\pm0.12$. These values were also calculated separately for each epoch, using the individual values of the spectral index, and then averaged. Errors of the averaged value consist of the standard deviation of averaging the ratios between epochs and the standard deviation  of averaging the component fluxes for each epoch, added in quadrature.
% $r_{\delta} = \delta_{o}/\delta_{i} = r_{I}^^{2+\alpha}$.

The apparent kinematics of the relativistic blazar jet are determined by four parameters: the apparent speed $\beta_{app}$, the Lorentz factor $\Gamma$, the beaming factor $\delta$ and the angle of the jet to the line of sight $\Theta$, any two of which could be used to find two others \citep[e.g.][]{cohen2007}. In the inner and outer parts of the jet, the Lorentz factor $\Gamma$ is the same, but the other three parameters differ: $\Theta_{2}>\Theta_{1}$, and therefore ${\beta_{app}}_{2}<{\beta_{app}}_{1}$ and ${\delta}_{2}<{\delta}_{1}$. From the 22 and 43 GHz data analysis \citep{rastorgueva2008} we estimated the kinematic parameters for the inner jet: the maximum apparent speed is ${\beta_{app}}_{max} = 19.4$ and the corresponding minimum Lorentz factor is $\Gamma_{min} = 19.5$. The corresponding minimum angle of the jet to the line of sight is $\Theta_{mins} = 3.0^{\circ}$.%, and the corresponding beaming factor in this %case would be $\delta =19.4$. At the same time, variability gives the beaming factor of $\delta_{var} = 10.9$, %which implies  whi 
%The fact that the bending of the jet causes simultaneous decrease of the both apparent speed and apparent flux suggests that the initial angle of the (inner) jet to the line of sight is larger than the $\Theta_{SL}$, but less than the maximal value of $\Theta_{max} = \arcsin(2 \beta_{app}/(1+\beta_{app}^2) = 5.9^{\circ}$ (all possible values lie in the lower part of a $\beta_{app} - \delta$ diagram). 
In the outer jet, the maximum apparent speed was $\beta_{app} \simeq 10$. Taking into account the above mentioned observed quantities and assuming that the jet has a constant speed, we can calculate its angle to the line of sight before ($\Theta_{1}$) and after ($\Theta_{2}$) the bending as well as other kinematic parameters of the jet. 

We can consider the beaming factor of the jet as a function of the initial (inner jet) angle to the line of sight $\Theta_{1}$, which is larger than the $\Theta_{min}=3.0^{\circ}$, but smaller than the maximal value of $\Theta_{max} = \arcsin(2 \beta_{app}/(1+\beta_{app}^2) = 5.9^{\circ}$ (all possible values lie in the lower part of a $\beta_{app} - \delta$ diagram). The ratio of the beaming factors $r_{\delta}$ is a monotonic function of $\Theta_{1}$ and changes from 0.15 to 0.26, while $\Theta_{1}$ changes within its limits, and reaches the value of 0.24 at $\Theta_{1} = 5.2^{\circ}$. Therefore, the observed change of kinematics is possible if the Lorentz factor of the jet is $\Gamma \simeq 30.0$, and at the apparent distance of 1~mas from the core the jet changes its direction, with the angle to to the line of sight increasing from $\sim 5^{\circ}$ to $\sim 11^{\circ}$. At the same time the beaming factor is changing from $\sim 7$ to $\sim 2$, respectively. 

The errors of the observed flux factor ratio determination, which give us a beaming factor ratio, are fairly large, and precise calculations of the angle to the line of sight and beaming factor are not possible. The values we obtained are only estimates, and we can only say that the observed change of the jet apparent speed and brightness is possibly caused by jet bending. Also, the proposed beaming factor for the inner jet is consistent with the value found by \citet{hovatta2009} from the radio-frequency variability. 

%\subsection{Helical jet model: application to the 5 GHz data}
%From the analysis of the 22 and 43~GHz data \citep[Paper I][]{rastorgueva2008} it was discovered that the components in the inner jet are likely to move along helical trajectories. It is possible to explain the 5~GHz kinematics with the same hypothesis. After 1~mas from the core, the observed jet becomes mostly diffuse, having only four components which apparently have chaotic motions on the small scale and rather stable flux densities and position (within errors). Three of them, C3, C4, and C5, are located about every 1.5 mas from the core and one (C6) at about 12 mas. An underlying helical jet with a large $\Gamma$ would be consistent with these observations: every $\sim$ 1.5 mas the local jet alignment is optimal and due to Doppler boosting the weak jet emission becomes visible. Then, after the third enhancement, the jet axis bends and the jet alignment angle stays large for the next four rotations with reduced boosting. At around 12 mas the jet axis bends again towards the line of sight and becomes visible. Unfortunately, the data we have are not enough to calculate the helix parameters: more sensitive VLBI observations with high resolution are needed for this task.
%
\section{Conclusions}
We studied the large-scale jet kinematics of the blazar S5 0716+714 by analyzing six epochs of VLBA observations at 5 GHz. Two imaging methods were applied to this data set: a conventional method (CLEAN self-calibration loop in 
\textit{Difmap} combined with \textit{Modelfit}) and GMEM (Pulkovo \textit{VLBImager} package combined with the image plane model-fitting).

\subsection{Performance of the methods}
S5 0716+714 has a bright point-like core and a faint diffuse jet. The emission fades down very quickly in the vicinity of the core, and the CLEAN algorithm, producing the model that is composed of the discrete point sources, is not adequately restoring this emission region. In the Gaussian model, some fraction of the core emission was added to the inner jet component, and the kinematic results were also ambiguous because of the large uncertainties of the positions of jet components.

The GMEM method delivered a smooth image of the source, which restored both the bright core and the diffuse jet. However, the jet of 0716+714 is very faint, therefore a convolution of the original solutions with the small (0.5$\times$0.5~mas) beam was needed in order to obtain accurate Gaussian models of the jet structure. The original GMEM solutions convolved with the CLEAN beam in general resemble the CLEAN images (Fig.~\ref{map:GMEMmaps} and \ref{map:CLEANmaps}).%, \ref{Appendix}~Appendix).

We recommend GMEM in combination with the difference mapping technique for restoring the structure of compact AGN and other objects that have a diffuse structure with bright point-like features.  In addition, the original GMEM solutions have a higher resolution than other methods, and in principle they could be used directly to derive kinematics of the source, provided that the jet features are bright enough \citep[e.g.,][]{bajkova2008}. 

\subsection{Structure and kinematics of the jet}
The large-scale jet of 0716+714 is faint, diffuse and smooth, without prominent brightness enhancements. 

Both the conventional method and GMEM deliver similar kinematic results: the large-scale jet (1-12~mas) is mostly stationary, with a large scatter of the component positions. 

The position angle of the jet on the sky plane changes by $\sim 10^{\circ}$ at about 1~mas from the core. 

Large-scale jet components are systematically slower and fainter than those of the inner ($<$1~mas) jet components. 

We attribute these last effects to the bending of the jet at the apparent distance $\sim$ 1~mas from the core from $\sim 5^{\circ}$ to $\sim 11^{\circ}$.

%9. The structure of the large-scale jet is likely to retain helical shape which was observed in the inner jet.

\subsection{Acknowledgements}
We thank the anonymous referee for the useful comments which have helped us to revise and improve the paper.

The National Radio Astronomy Observatory is a facility of the National Science Foundation operated under cooperative agreement by Associated Universities, Inc. 

This study was supported in part by the ``Origin and Evolution of
Stars and Galaxies'' \-- ``Program of the Presidium of the Russian
Academy of Sciences and the Program for State Support of Leading
Scientific Schools of Russia'' (NSh--3645.2010.2).

\bibliographystyle{aa} % style aa.bst
\bibliography{0716_refs3} % your references Yourfile.bib

\newpage
\onecolumn
\section{Appendix} \label{Appendix}


\begin{table*}
\caption[]{Results from the CLEAN and Gaussian model fitting and
component parameters. $S_{peak}$ is the peak flux density, r and
PA are the distance and position angle measured from the core and
$a_{gauss}$ is the FWHM of the circular Gaussian component. ID is a component
identification in one of the three possible scenarios,
see Section~\ref{Results}.}
\label{tb:CLEAN_models} \centering
\begin{tabular}{clrrll}
\hline\hline

Epoch, & Flux $I$       & r,  & PA,        & $a_{gauss}$,  & ID\\
 yr   & [mJy] & [mas] & [$^{\circ}$] & mas         & \\
\hline
2004.11 & 1282.0$\pm$10.0 & 0~~~~~~~~~~~~~~~ & 0~~~~~~~~~~~ & 0.05   & Core \\
    & ~~138.0$\pm$10.0 & 0.40$\pm$0.03 & 27.0$\pm$0.5 & 0.13   & C1-2\\
    & ~~~~14.6$\pm$3.0 & 1.64$\pm$0.13 & 22.0$\pm$1.0 & 0.59   & C3 \\
    & ~~~~~~8.5$\pm$3.0 & 3.43$\pm$0.11 & 16.5$\pm$0.3 & 0.96   & C4\\
    & ~~~~~~4.2$\pm$1.0 & 6.38$\pm$0.21 & 10.0$\pm$1.2 & 2.87   & C5 \\
    & ~~~~~~8.0$\pm$2.0 & 11.59$\pm$0.27 & 18.0$\pm$0.3 & 3.76   & C6\\
2004.22 & ~~989.0$\pm$10.0 & 0~~~~~~~~~~~~~~~ &   0~~~~~~~~~~~   &  0.08 & Core \\
    & ~~114.0$\pm$23.0 & 0.58$\pm$0.02 & 26.0$\pm$0.3  &  0.14   & C1-2 \\
    & ~~~~~~9.2$\pm$1.0 & 2.48$\pm$0.11 & 20.0$\pm$1.2 & 0.59   & C4\\
    & ~~~~~~6.6$\pm$1.0 & 4.89$\pm$0.33 & 13.8$\pm$2.4 & 2.11   & C5\\
    & ~~~~~~6.4$\pm$1.0 & 11.35$\pm$0.33 & 16.0$\pm$1.1 & 2.80   & C6\\
 2004.34 & ~~908.0$\pm$2.0 & 0~~~~~~~~~~~~~~~ & 0~~~~~~~~~~~ & 0.07  & Core \\
     & ~~121.0$\pm$3.0 & 0.60$\pm$0.02 & 26.0$\pm$0.4 & 0.20   &C1-2 \\
    & ~~~~~~8.6$\pm$3.0 & 1.60$\pm$0.22 & 20.0$\pm$1.3 & 9.0e-06   & C3\\
    & ~~~~10.7$\pm$0.4 & 3.20$\pm$0.07 & 16.1$\pm$0.3 & 0.97   & C4 \\
    & ~~~~~~5.4$\pm$1.0 & 6.90$\pm$0.20 & 12.0$\pm$1.0 & 2.96   & C5 \\
    & ~~~~~~7.0$\pm$1.0 & 12.07$\pm$0.24 & 18.7$\pm$0.7 & 4.22   & C6\\
2004.46 & ~~920.0$\pm$60.0 & 0~~~~~~~~~~~~~~~ & 0~~~~~~~~~~~ & 0.08   & Core \\
    & ~~~~90.4$\pm$50.0 & 0.28$\pm$0.08 & 58.0$\pm$7.0 & 1.3e-04   & X \\
    & ~~114.0$\pm$40.0 & 0.72$\pm$0.05 & 25.0$\pm$1.7 & 0.19   & C1-2 \\
    & ~~~~~~9.4$\pm$3.0 & 1.97$\pm$0.25 & 22.0$\pm$1.7 & 0.77   & C3\\
    & ~~~~~~6.5$\pm$2.0 & 3.56$\pm$0.20 & 14.7$\pm$1.3 & 0.78   & C4 \\
    & ~~~~~~4.5$\pm$1.0 & 5.56$\pm$0.29 & 18.3$\pm$3.0 & 2.50   & C5 \\
    & ~~~~~~6.7$\pm$13.0 & 11.34$\pm$0.28 & 16.0$\pm$0.3 & 3.33   & C6 \\
2004.58 & ~~724.0$\pm$7.0 & 0~~~~~~~~~~~~~~~ & 0~~~~~~~~~~~ & 2.4e-3   & Core \\
    & ~~143.0$\pm$7.0 & 0.68$\pm$0.015 & 29.5$\pm$0.7 & 0.24   & C1-2 \\
    & ~~~~~~6.3$\pm$2.0 & 1.77$\pm$0.26 & 21.7$\pm$2.7 & 3.0e-07   & C3\\
    & ~~~~~~8.2$\pm$1.0 & 3.16$\pm$0.20 & 21.3$\pm$2.3 & 0.91   & C4\\
    & ~~~~~~5.8$\pm$1.0 & 5.09$\pm$0.40 & 11.7$\pm$2.0 & 1.61   & C5\\
    & ~~~~~~9.6$\pm$2.0 & 10.85$\pm$0.37 & 16.7$\pm$2.7 & 4.58   & C6 \\
2004.66 & ~~738.0$\pm$5.0 & 0~~~~~~~~~~~~~~~ & 0~~~~~~~~~~~ & 0.06   & Core \\
    & ~~104.0$\pm$4.0 & 0.81$\pm$0.02 & 27.7$\pm$0.8 & 0.12  & C1-2 \\
    & ~~~~~~6.5$\pm$2.0 & 2.43$\pm$0.50 & 18.7$\pm$3.0 & 0.69   & C3\\
    & ~~~~~~6.4$\pm$2.0 & 4.21$\pm$0.60 & 17.0$\pm$7.0 & 1.16   & C4\\
    & ~~~~~~8.7$\pm$2.0 & 8.71$\pm$0.48 & 11.7$\pm$3.3 & 4.36   & C5 \\
    & ~~~~~~2.1$\pm$2.0 & 12.31$\pm$0.50 & 24.3$\pm$2.3 & 1.25  & C6 \\

 \hline
\end{tabular}
\end{table*}
\begin{table*}
\caption[]{Results from the GMEM method imaging and model fitting with
component parameters. $S_{peak}$ is the peak flux density, r and
PA are the distance and position angle measured from the core, $a_{major}$ and $a_{minor}$ are the major and minor axes of the elliptical Gaussian components and $\theta$ is a position angle of the ellipse. ID is a component
identification in the stable jet scenario,
see Section~\ref{Results}. }
\label{tb:GMEM_models} \centering %\begin{small}
\begin{tabular}{clrlccrl}
\hline\hline
Epoch & Flux $I$        &  $r$ & $ PA $ & $a_{major}$ & $a_{minor}$ & $\theta$ & ID \\
      & [mJy]             & [mas]& [$^{\circ}$] & [mas] &  [mas]   & [$^{\circ}$] &   \\
\hline
2004.11&1414.5       &  0~~~~~~~~~~~~~~~~~~~   & ~~0    & 0.61 & 0.54 & 8.99   & Core\\
        &~~~~~~7.4   & 0.486$\pm0.006$ &28.56 & 0.61 & 0.51 & 138.14 & A1\\
        &~~~~~12.1   & 1.664$\pm0.007$ &31.18 & 0.75 & 0.66 & -30.33 & A3\\
        &~~~~~~3.0   & 2.502$\pm0.013$  & 5.24 &1.22  & 0.76 &  24.61 & X \\
        &~~~~~~6.1   & 3.470$\pm0.043$ &13.44 &1.20  & 0.83 & 117.53 & A4\\
        &~~~~~~7.0   & 5.398$\pm0.342$ &12.72 &2.37  & 1.94 & 279.91 & A5\\
        &~~~~~~7.7   & 12.542 $\pm$0.679& 15.29& 5.57& 1.39& 53.18&    A6\\

2004.22&1011.8 & 0~~~~~~~~~~~~~~~~~~~    & ~~0    &0.65  & 0.55 & 203.02 & Core \\
        &~~~~19.9  & 0.486$\pm0.002$ &22.08 &0.64  & 0.57 & 33.93  &  A1\\
        & ~~~~~14.9  & 1.533$\pm0.014$ &25.75 &1.10  & 0.89 & 45.55  &  A3\\
        & ~~~~~~8.9  & 2.984$\pm0.060$ &14.01 &1.71  & 0.92 & 98.89  &  A4\\
        & ~~~~~~3.8  & 4.635$\pm0.140$ &15.09 &1.71  & 0.90 &103.87  &  A5\\
        &~~~~~~3.5   & 11.156 $\pm$ 0.021 & 8.65 & 0.99 & 0.62 & 24.58  &A6\\

2004.34&~~945.6 & 0~~~~~~~~~~~~~~~~~~~   & ~~0     &0.64  & 0.55 & 21.41  & Core\\
        &~~~~28.7 & 0.462$\pm0.001$ & 26.63 &0.59  & 0.52 &-60.54  & A1\\
        &~~~~19.4  & 0.986$\pm0.004$ & 28.59 &0.87  & 0.60 &34.05   & A2\\
        &~~~~~~8.5  & 1.483$\pm0.016$ & 24.74 &1.15  & 0.66 &145.82  & A3\\
        &~~~~~~8.5  & 3.315$\pm0.022$ & 13.14 &1.14  & 0.92 &-202.12 & A4\\
        &~~~~~~3.2  & 4.775$\pm0.110$ & 19.86 &1.69  & 0.77 & 41.74  & A5\\
        &~~~~~~3.2  & 7.532$\pm0.292$ & 19.04 &2.85  & 0.71 & 25.94  & X\\
        &~~~~~~5.8  & 13.10 $\pm$  0.666        & 18.03 &4.99  & 1.31 & 38.75 &
        A6\\
2004.46&~~961.0& 0~~~~~~~~~~~~~~~~~~~    & ~~0     &0.65  & 0.55 & 21.80  & Core\\
        &~~~~29.1 & 0.489$\pm0.001$ &26.83  &0.56  & 0.53 &112.24  & A1\\
        &~~~~32.5 & 0.916$\pm0.003$ &27.80 &0.96  & 0.61 &-144.55 & A2\\
        &~~~~~~6.6  & 1.549$\pm0.038$ &16.70  &1.39  & 0.69 &134.55  & A3\\
        &~~~~~~6.1  & 3.399$\pm0.042$ &14.19  &1.30  & 0.81 & -170.79& A4\\
        &~~~~~~3.5  & 4.601$\pm0.240$&20.80    &2.21  &0.91  &118.47 & A5\\
        &~~~~11.8  & 11.11  1.711        &16.25 & 8.39& 2.36& 159.12& A6\\
2004.58&~~711.7 & 0~~~~~~~~~~~~~~~~~~~   & ~~0     &0.66  & 0.56 & 24.10  & Core\\
        &~~~~36.0  & 0.489$\pm0.001$& 25.70 &0.57  & 0.56 &90.70   & A1\\
        &~~~~38.5  & 0.978$\pm0.001$& 27.75 &0.86  & 0.60 &28.13   & A2\\
        &~~~~~~6.9   & 1.360$\pm0.009$& 17.91 &0.94  & 0.59 &6.29    & A3\\
        &~~~~~~7.1   & 3.367$\pm0.025$& 15.00 &1.29  & 0.88 &-25.33  & A4\\
        &~~~~~~4.5   & 4.527$\pm0.141$& 22.39 &2.22  & 1.09 &115.54  & A5\\
        &~~~~~13.3   & 11.33 $\pm$ 2.216        & 18.12 &8.45  & 3.61 &203.37
        &A6\\
2004.66&~~712.2 & 0~~~~~~~~~~~~~~~~~~~   & ~~0     &0.70  & 0.56 & 19.46  & Core\\
        &~~~~27.4  & 0.523$\pm0.001$& 21.35 &0.60  &0.55  & 98.63  & A1\\
        &~~~~~~7.9   & 0.852$\pm0.003$& 21.66 &0.57  &0.53  & 38.70  & A2\\
        &~~~~25.8  & 1.300$\pm0.003$& 25.19 &0.91  &0.66  &-150.48 & A3\\
        &~~~~~~4.8   & 3.558$\pm0.108$& 16.51 &2.07  &1.02  &-52.99  & A4\\
        &~~~~~~4.8   &4.189$\pm$0.083 & 15.81 &1.92  &0.89  & 33.07  & A5\\
        &~~~~~~1.4   &6.886$\pm$0.062 & 19.18 &1.06  &0.62  &196.45  &  X \\
        &~~~~~~9.3   & 12.40$\pm$0.794 & 12.37 & 6.83& 1.31  & 59.65 & A6\\
\hline
\end{tabular} %\end{small}
\end{table*}

\end{document}